\documentclass[journal]{IEEEtran}
\usepackage{amsthm}
\usepackage{cite}
\usepackage{amsmath}
\usepackage{nccmath}
\usepackage{amssymb}
\usepackage{amsfonts}
\usepackage{graphicx}
\usepackage{fancyhdr}
\usepackage{svg}
\usepackage{bbm}
\usepackage{subcaption}
\usepackage{textcomp}
\usepackage{xcolor}
\usepackage[linesnumbered,ruled]{algorithm2e}
\usepackage{booktabs}
\usepackage{bbold}
\usepackage{kotex}

\theoremstyle{definition}

\usepackage{multicol}
\usepackage{multirow}

\usepackage[normalem]{ulem}
\usepackage{lettrine}

\newcommand{\BfPara}[1]{{\noindent\bf#1.}\xspace}

\begin{document}

\title{Age-of-Information Aware Contents Caching and Distribution for Connected Vehicles}

\author{
    Soohyun Park,
    Chanyoung Park, 
    Soyi Jung,~\IEEEmembership{Member, IEEE},
    Minseok Choi,~\IEEEmembership{Member, IEEE},
    and \\
    Joongheon Kim,~\IEEEmembership{Senior Member, IEEE}
    \thanks{This research was funded by National Research Foundation of Korea (2022R1A2C2004869, 2021R1A4A1030775). \textit{(Corresponding authors: Soyi Jung, Minseok Choi, Joongheon Kim)}}
    \thanks{Soohyun Park, Chanyoung Park, and Joongheon Kim are with the School of Electrical Engineering, Korea University, Seoul 02841, Republic of Korea (e-mails: \{soohyun828,cosdeneb,joongheon\}@korea.ac.kr).}
    \thanks{Soyi Jung is with the Department of Electrical of Computer Engineering, Ajou University, Suwon, Republic of Korea (e-mail: sjung@ajou.ac.kr).}
    \thanks{Minseok Choi is with the Department of Electronic Engineering, Kyung Hee University, Yongin, Republic of Korea (e-mail: choims@khu.ac.kr).}
}
\maketitle

\begin{abstract}
To support rapid and accurate autonomous driving services, road environment information, which is difficult to obtain through vehicle sensors themselves, is collected and utilized through communication with surrounding infrastructure in connected vehicle networks. 
For this reason, we consider a scenario that utilizes infrastructure such as road side units (RSUs) and macro base station (MBS) in situations where caching of road environment information is required. 
Due to the rapidly changed road environment, a concept which represents a freshness of the road content, age of information (AoI), is important. Based on the AoI value, in the connected vehicle system, it is essential to keep appropriate content in the RSUs in advance, update it before the content is expired, and send the content to the vehicles which want to use it. However, too frequent content transmission for the minimum AoI leads to indiscriminate use of network resources. Furthermore, a transmission control, that content AoI and service delay are not properly considered adversely, affects user service. Therefore, it is important to find an appropriate compromise. For these reasons, the objective of this paper is about to reduce the system cost used for content delivery through the proposed system while minimizing the content AoI presented in MBS, RSUs and UVs. The transmission process, which is able to be divided into two states, i.e., content caching and service, is approached using Markov decision process (MDP) and Lyapunov optimization framework, respectively, which guarantee optimal solutions, as verified via data-intensive performance evaluation. 
\end{abstract}

\begin{IEEEkeywords}
Caching system, age-of-information, 6G, Markov decision process, Lyapunov optimization.
\end{IEEEkeywords}
\IEEEpeerreviewmaketitle

\section{Introduction}\label{sec:introduction}
\subsection{Backgrounds and Motivation}
\IEEEPARstart{S}{mart} vehicles that intelligently assist drivers or have advanced autonomous driving technologies interact with their surroundings in real-time as well as determine optimal driving decisions for safe and fast driving~\cite{ijcai2019shin,pieee202105park}. 
For the purpose of rapid and accurate driving decisions to ensure the driving stability in fast-moving connected vehicle network environment, related studies have attracted explosive attention. Until now, research on optimal driving policy-making algorithms using reinforcement learning or data transmission algorithms using unmanned autonomous vehicles (UAVs) and surrounding infrastructure that efficiently delivers road environment to vehicles have been actively conducted~\cite{tvt202108jung,jcn2022haemin,9467353,tvt202106jung,tvt201905shin}. The technologies in various fields are being studied to advance the connected vehicle technology. Among them, rapid data delivery and sharing using the infrastructure of the vehicle network is especially important because it is used as the basis for control decisions through driving policies~\cite{IoTJ_DRL_caching}.
The connected smart vehicles can collect environmental information and vehicle condition using various built-in sensors. In addition, they share and collect necessary information through connection with infrastructure such as road side units (RSUs) based on Internet-of-things (IoTs) or internet-of-vehicles (IoVs) technologies~\cite{9194445}. The collected data quality (e.g., image quality, amount of information contained, data oldness, and data suitability of purpose, etc.) partially affects the driving stability of the vehicle. For this reason, we are interested in how to efficiently support road content that contains environmental information and design appropriate solutions for the connected vehicles~\cite{tvt202108jung}.
Here, we have to keep in mind that external information that vehicles cannot obtain through internal sensors can be obtained through other media such as nearby vehicles, RSUs, and drones. However, when the vehicle is far from a target which generates road contents and belongs to the other area where communication is impossible, the data transmission will be interrupted and the vehicle can not receive the necessary data on time. For this reason, content management and service with storage which merges all the data that comes from the network is important.

However, the aggregation of all data in the center of the network causes an unexpected delay in providing the requested content and unnecessary waste of communication cost and server storage~\cite{JSAC_caching_in_MEC_RL}. 
As a way to solve this problem, the vehicle networks can use distributed sub-storage (e.g., RSUs with cache) connected to the central base station. By distributing some contents in each RSU, it is possible to deliver necessary information to the data requestors nearby. Especially, research using the distributed cache has been focused on streaming applications~\cite{TWC_MDP_cahcing, TWC_distributed_caching, TWC_probabilistic_caching, IoTJ_on-demand_streaming_IoV}.
The distributed cache dramatically reduces the system backhaul cost and transmission time used for content delivery from the central base station. Since the distributed cache is generally smaller than the central base station, it has limitations in terms of storage capacity to hold all contents the same as the central base station. Therefore, cache management considering these characteristics is essential. Furthermore, in resent years, the distributed cache concept is combined with vehicles or unmanned mobile objects such as UAVs and high altitude platforms (HAPs) to enable more adaptive and flexible response to the cached content requests~\cite{9687261}. However, if there are no repetitive features or specific patterns in the movement, instability can be a problem in the communication due to the activity of the mobile cache, such as time-varying content popularity, dynamic network topology, and vehicle driving path. These problems still remain to be solved in the use of mobile cache~\cite{IoTJ_DRL_caching}.

There are additional factors to consider in recent caching studies. If time flow and data characteristics or values are irrelevant, such as streaming using platforms (e.g., YouTube and Netflix), cache management is determined by the average popularity or the temporary surge of the contents. %~\cite{}.popularity기반 ref 추가 
However, as already mentioned, the vehicle network considered in this study requires the appropriate use of environmental information that changes over time. Because of this, unlike general cache management in video streaming applications, the freshness index of the data, which is called \textit{age-of-information (AoI)} should be considered~\cite{MILCOM_AoI}. AoI, a matrix that evaluates the freshness of data, is a value accumulated over time since the data is created. The increase of the value means that the data is generated a long time ago, and it can be considered that the effectiveness of the data decreases. In other words, The large value of AoI means that the data is too old to reflect the current environmental condition. For this reason, cache management which considers only the popularity of the data or the capacity of the cache causes fatal defects in vehicles that need to make stable driving decisions in real-time. Similarly, considering only AoI minimization overshadows the reduction of system communication cost, which is one of the purposes of using distributed caches. This results in excessive data exchange between MBS and RSU and a waste of communication resources. For these reasons, we are interested in AoI-based caching and user services utilizing the infrastructure of the vehicle network.

\subsection{AoI-Aware Algorithm Design Rationale}

\begin{figure}[t!]
    \begin{center}
        \includegraphics[width=1\linewidth]{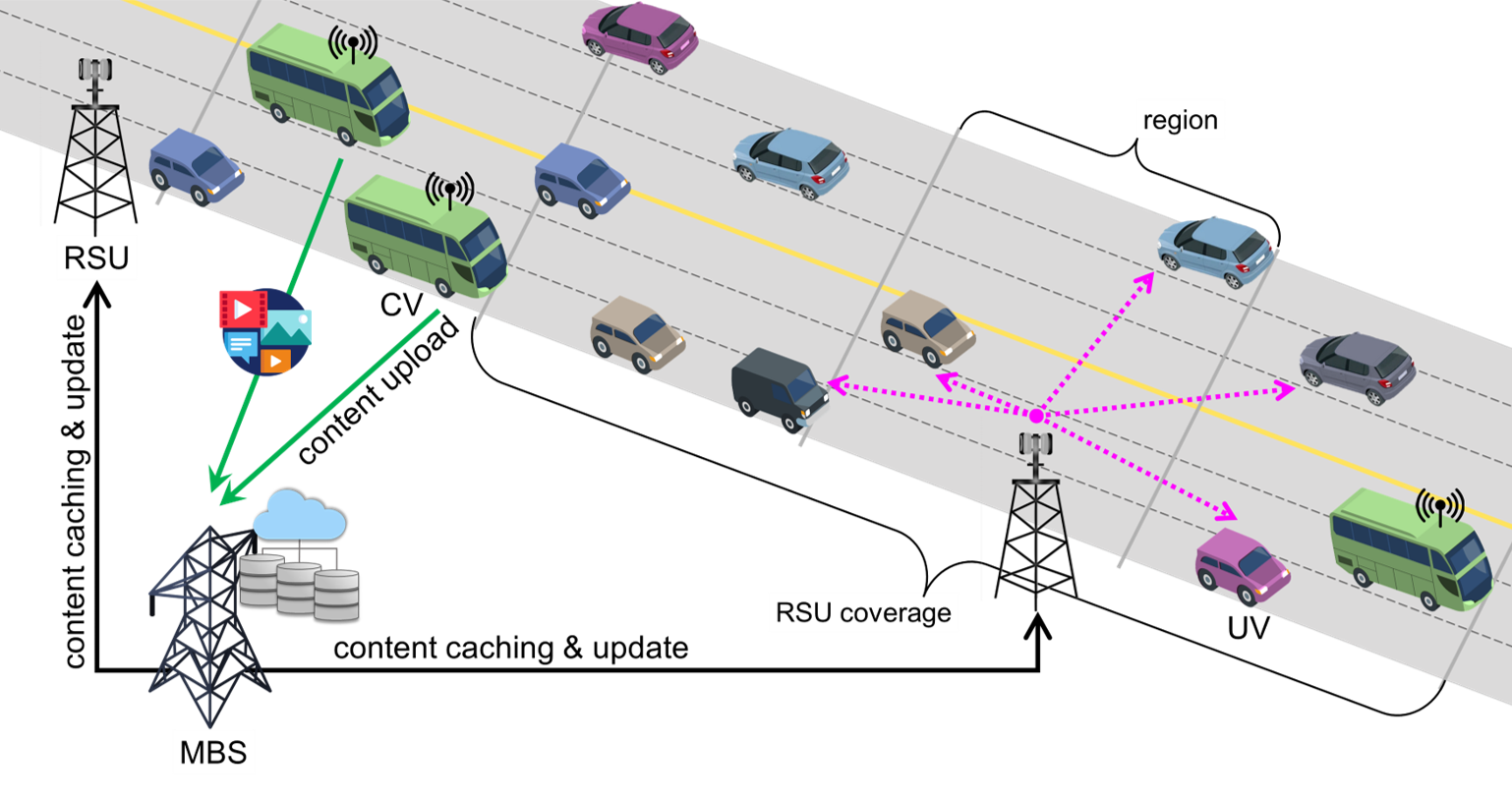}
    \end{center}
    \caption{Illustration of our considering scenario.}
    \label{network architecture}
\end{figure}

Based on the need for the above-distributed cache and the need to consider the AoI of the data stored in the cache, we present a new distributed caching network for connected vehicles. There are two types of vehicles in the proposed network: a connected user vehicle (UV) that requires road environment information for stable driving and a content producer vehicle (CV) that collects data and makes it as a content. CV does not play a role as a vehicle cache but collects road data while driving and delivers it to the network-centered base station (MBS). Unlike vehicle cache, there is no need to seriously consider vehicle storage capacity because all data sent from CV to MBS are erased. Content generated in CV is uploaded to MBS, cached into each RSU that requires the content, and finally delivered to UV through the RSU. In this case, in order to support fresh data to UVs, the AoI of the content flowing into the RSU through CV and MBS must have a sufficiently small value. Based on the AoI value of the content, the MBS brings the content from the CVs and updates the content cached in the RSU. For this reason, resource-effective content upload and update decisions which minimize content AoI considering communication costs is essential. In the proposed process, the CVs upload contents to MBS and MBS updates each RSU cache to recent data.

In addition, RSU, which has updated data with the latest data through MBS, directly serves UVs. For the UV service step, due to the vehicle mobility, fast data transmission is required, and the waiting time of UV should also be considered with AoI and communication cost. The UV sends the request for the target area content while driving and accumulates waiting time after requesting to the RSU. In this case, the UV should receive the desired content from the RSU within the time required for the content before passing through the target area so that the UV receives valid service. There is a limit to the number of the available channels that the RSU can simultaneously use, the state in which the service can wait, and the desired content are different for each UV. For this reason, UV's service delay and AoI are used as a factor of judgment in serving content for stable driving of UVs. In particular, the optimal allocation of RSU communication channels based on this is considered an important issue. That is, service control of RSU based on the request latency of connected vehicles along with content caching in a connected vehicle environment is also required for efficient operation of the entire system.

We approach the fresh data support problem that must be solved for stable and effective driving of the connected vehicle by dividing it into two stages: caching and service. The caching and content service processes are conducted independently of each other to minimize content AoI for each stage by consuming least cost. Content caching should complement the limitations of central storage and vehicle cache and allow content present in the system to reflect recent road conditions. The content service should ensure that the AoI of the content currently in the RSU is transmitted to the UV while ensuring its validity as data. We solve the content caching problem using MDP, which can always obtain optimal solutions and we use Lyapunov control to find the optimal solution considering AoI for service delay and communication cost in a trade-off relationship.

\subsection{Contributions}
The main contributions of this research are as follows:
\begin{itemize}
    \item We propose a new vehicle network architecture which is constructed by MBS, RSUs, CVs, and UVs. The road covered by RSUs is divided into several regions. Each region has a different road state and traffic condition. The road content is getting older after being generated by CVs. In the proposed connected vehicle network, data delivery through the road infrastructure (e.g., MBS, RSUs and CVs) reflecting rapidly changing road environment information is essential. For this reason, new caching research which is appropriate for the proposed network considering AoI is important.
    \item We propose optimal cache management and transmission decisions considering AoI which represents data freshness. In order to optimize the individual two decisions, we consider content AoI, communication cost, and waiting times. There are few studies on the problem of vehicle network caching considering the three factors simultaneously. 
    \item We make an approach which divides the transmission process occurring in the network into two stages: cache management (upload and update) and content service to establish an object for each stage and optimize it independently. Each of the two stages uses MDP and Lyapunov control, which always guarantees optimalit~\cite{ton201608kim,isj202108jung}. Although it is not a joint optimal relationship whith each other, the purpose of the two stages' combination is to ensure that UV receives the latest information and drives stably. 
\end{itemize}

\begin{table}[t]
    \caption{Key Notations}
    \label{tab:notation}
    \centering
    \begin{tabular}{p{1.2cm}||p{6cm}}
        \toprule[1.0pt]
        Notation & Description\\
        \midrule[1.0pt]
        $N_u$ & Index set of user vehicle (UV)\\
        $N_c$ & Index set of content producer vehicle (CV)\\
        $N_R$ & Index set of content caching RSU\\
        $L$ & Index set of road region\\
        $V^{u}_i$ & $i$-th UV $i \in N_u$\\
        $V^{c}_j$ & $j$-th CV $j \in N_c$ \\
        $R_k$ & $k$-th RSU $k \in N_R$\\
        $C^{c}_{j,h}$ & Content of region $h\in L$ collected by $j$-th CV\\
        $C^{R}_{k,h}$ & Content of $h$-th region cached in RSU $k \in R_k$\\
        $C_{h}$ & Content of $h$-th region stored in MBS\\
        $A^{c}_{j,h}$ & AoI value of content of region $h$ in $j$-th CV\\
        $A^{R}_{k,h}$ & AoI value of content of region $h$ cached in $k$-th RSU\\
        $A_{h}$ & AoI value of content of $h$-th region stored in MBS\\
        $A^{max}_h$ & AoI maximum value for content of region $h \in L$\\
        \bottomrule[1.0pt]
    \end{tabular}
\end{table}

\subsection{Organization and Key Notations}
The remainder of the paper is organized as follows. Sec.~\ref{sec:related_work} presents the related works followed by the system architecture and problem definition described in Sec.~\ref{sec:system_model}. Sec.~\ref{sec:proposed1} and Sec.~\ref{sec:proposed2} present our proposed optimal contents caching and UV content service algorithm. Sec.~\ref{sec:simulation} presents the realistic simulation parameters and analyzes the evaluation results. Finally, Sec.~\ref{sec:conclusion} concludes this paper. The key notations of this paper is listed in Tab.~\ref{tab:notation}.

\section{Related Work}\label{sec:related_work}
\subsection{Caching in Connected Vehicle Networks}
% content popularity based classical caching 
A popularity-based caching strategy has been researched as a solution for the finite-buffer restrictions in the traditional caching system.
Due to buffer capacity limitations, only the content file with the highest level of popularity may be cached. Deterministic and random caching strategies were researched to improve the cache hit ratio~\cite{TWC[11]} and decrease latency~\cite{TWC[12]} with finite buffers under the presumption that content popularity was known or totally predictable. The popularity of a piece of material can be forecasted in situations when the popularity profiles are unknown using the request history. In~\cite{TWC[14]}, a caching method with popularity prediction is suggested. The spatial and temporal differences among users' preferences were considered in~\cite{TWC[15]}.  
An major issue with the caching system is how to predict time-varying content popularity in actuality. A research of online caching using information theory is suggested, in~\cite{TWC[16]}
There are learning-based methods for reliably predicting content popularity. According to~\cite{TWC[17]}, an online proactive caching system that is based on a recurrent neural network model and can monitor popularity over time is offered. In~\cite{TWC[21], TWC[22], TWC[23]}, the problem is formulated using MDP. Particularly in~\cite{TWC[23]}, the reinforcement learning method achieves the long-term average energy cost reduction while preserving the cache threshold.
In~\cite{TWC_MDP_cahcing}, Utilizing request delay information, or the forecast of the user's request time, caching rules with limited buffers can increase the cache hit ratio. In the system, there is a data link  that connects the users and the BS, and during each time slot, the BS may actively push these content files to the user. To achieve the goal of increasing the average cache hit ratio, the MDP technique is used to tackle the issue.

% caching in vehicular network 
In vehicular networks, most studied caching schemes focus on caching at the supporting infrastructure (e.g., RSUs). In~\cite{TVT[23]} and~\cite{TVT[24]}, content downloading delay minimization by optimal placement scheme of popular files at RSUs is considered. Additionally, a competing content provider auction-based approach is suggested~\cite{TVT[25]}. In~\cite{TVT[26]}, the goal of heterogeneous vehicular networks with macro base stations and cache-enabled green RSUs is to reduce the cost of network construction while taking into account backhaul capacity and requirements for quality of service. In~\cite{TVT[27], TVT_IV-cache_Vehicular_network}, in-vehicle caching is proposed. Especially, in~\cite{TVT_IV-cache_Vehicular_network}, to transport the stored data from the leaving vehicle to the other vehicles through one-hop V2V networks, they allot data transfer areas.

\subsection{AoI-based Content Transmission}
AoI is a metric for information freshness that measures the time that elapses since the last received fresh update was generated at the source~\cite{TIT[11], TIT[12]}. The AoI increases linearly in time, until the destination receives a fresh update packet.  
Minimizing AoI, which means oldness in data, is a study that has received a lot of attention~\cite{TIT[1], TIT[35]}. 
In an environment where data updates are required (e.g. mobile device's recent position, speed, and other control information), the analysis and optimization of the AoI performance have been extensively studied in various scenarios~\cite{TIT[8], TIT[9], TIT[10], TIT_AoI}. 
In particular, in applications such as ultra-reliable vehicular communication~\cite{TCOM[14], TCOM[15]}, random access~\cite{TCOM[16]} and caching replacement~\cite{TCOM[17], TCOM[18], TCOM[19]}, AoI is used as an important evaluation index~\cite{TCOM_AoI_caching}.  
Since AoI is in a trade-off relationship with communication cost, transmission delay, and cache capacity, which are naturally important in caching systems, most studies consider different values along with AoI~\cite{TIT_AoI}. 
In~\cite{TWC_AoI_caching}, presents a content refresh algorithm for a mobile edge-based caching system to balance service latency and content freshness determined by AoI. The authors define a refresh window as the threshold AoI value at which a piece of material is considered valuable. The system only updates the material when the AoI goes above the threshold. AoI and delay have a trade-off connection with regard to the refreshing window. To minimize average latency and yet satisfy AoI criteria, the window size can be optimized.
In a mobile edge computing environment where information delivered from the sensor is processed and stored at the distributed edge, reducing the computing offloading cost of mobile user maintaining the freshness of contents cached in edges is proposed~\cite{IoTJ_freshness}. Due to the restricted wireless bandwidth of edges, it is vital to consider communication costs. The channel allocation and compute offloading control have been concurrently optimized for the first time to lower the overall cost while maintaining the required freshness. By converting the AoI recursiveness into a queue evaluation, the suggested technique optimizes the two values in the trade-off relationship based on Lyapunov optimization.

\section{System Model}\label{sec:system_model}
In this section, we describe the system model to which the proposed problem and solution are applied.
Through the following two subsections, we explain the network model, defined AoI concept, and the problem formulation which has to be solved.

\subsection{Distributed Connected Vehicle Networks}\label{sec:system model-network}
In the proposed network, we consider that content transmission which contains (i) content upload between CVs and MBS, (ii) content update between MBS and RSUs, and (iii) content service from RSUs to UVs is achieved in one time slot independently. In addition, we assume that there are no transmission failure factors such as packet loss and interference after deciding on the transmission for the three cases.

\subsubsection{Network Model}

\begin{figure}[t]
    \begin{center}
        \includegraphics[width=0.95\linewidth]{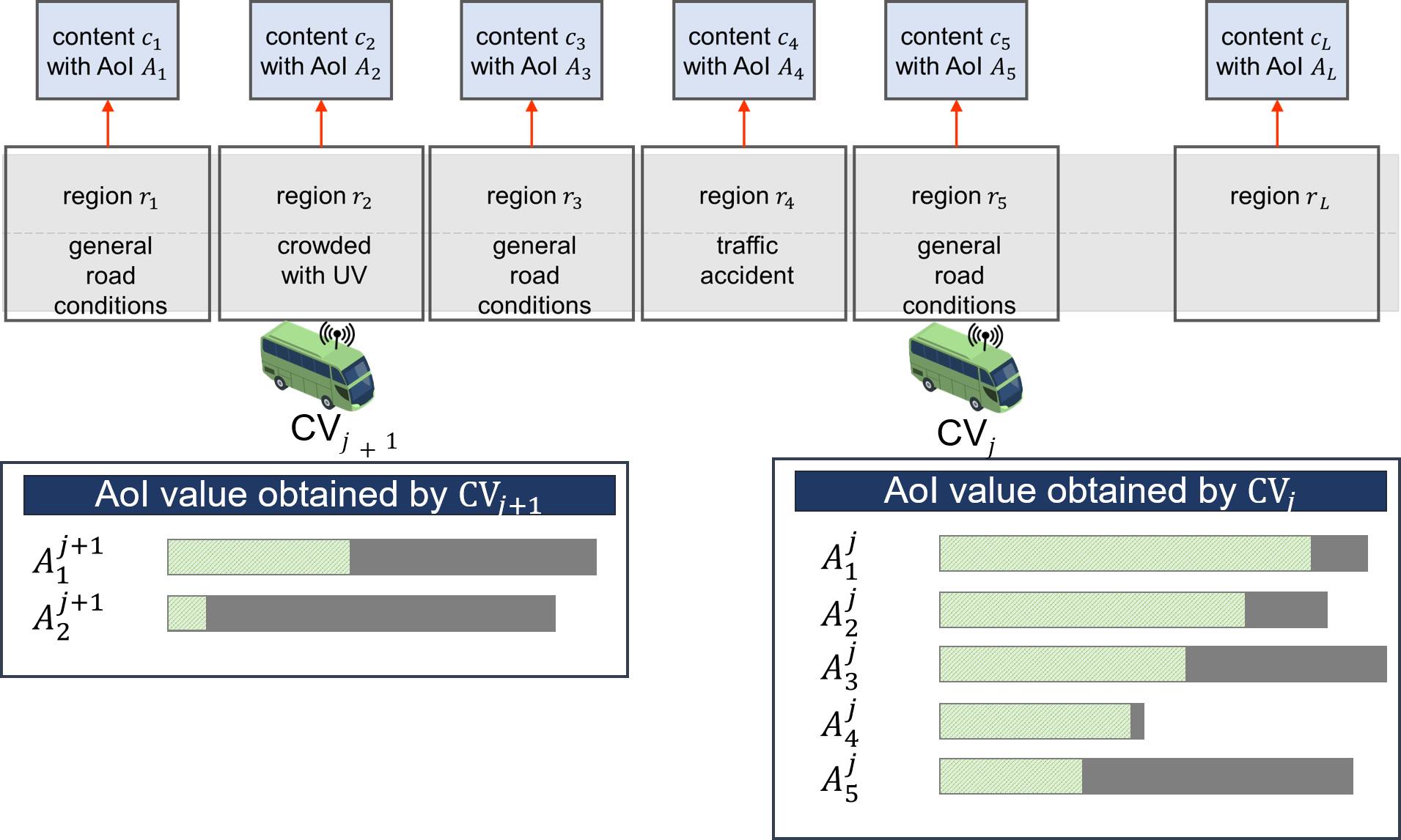}
    \end{center}
    \caption{Content AoI of CV for the road region.
    For the region which has different road condition, the content AoI maximum value $A^{max}_h$ is set depending on the condition to reflect the latest road conditions as well as possible. The CV which path through the region long ago has more older contents than the CV that comes later.}
    \label{fig:CV}
\end{figure}

\begin{figure}[t]
    \begin{center}
        \includegraphics[width=0.95\linewidth]{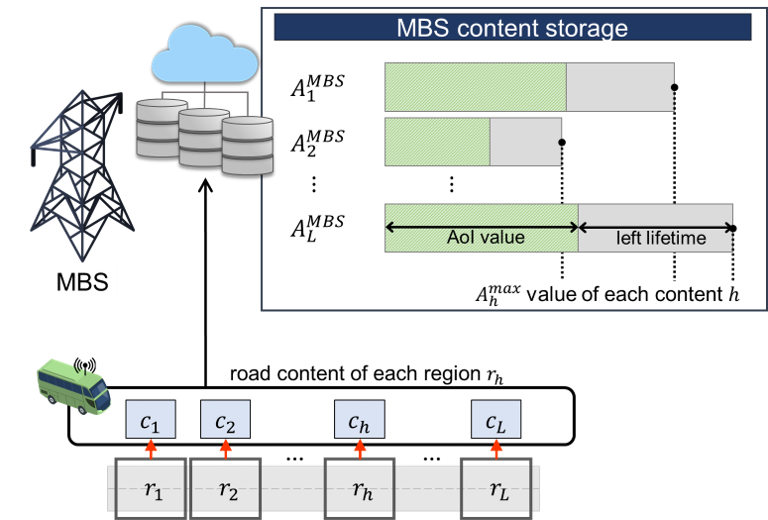}
    \end{center}
    \caption{Content uploading between MBS and CVs.}
    \label{fig:MBS}
\end{figure}

\begin{figure}[t]
    \begin{center}
        \includegraphics[width=0.95\linewidth]{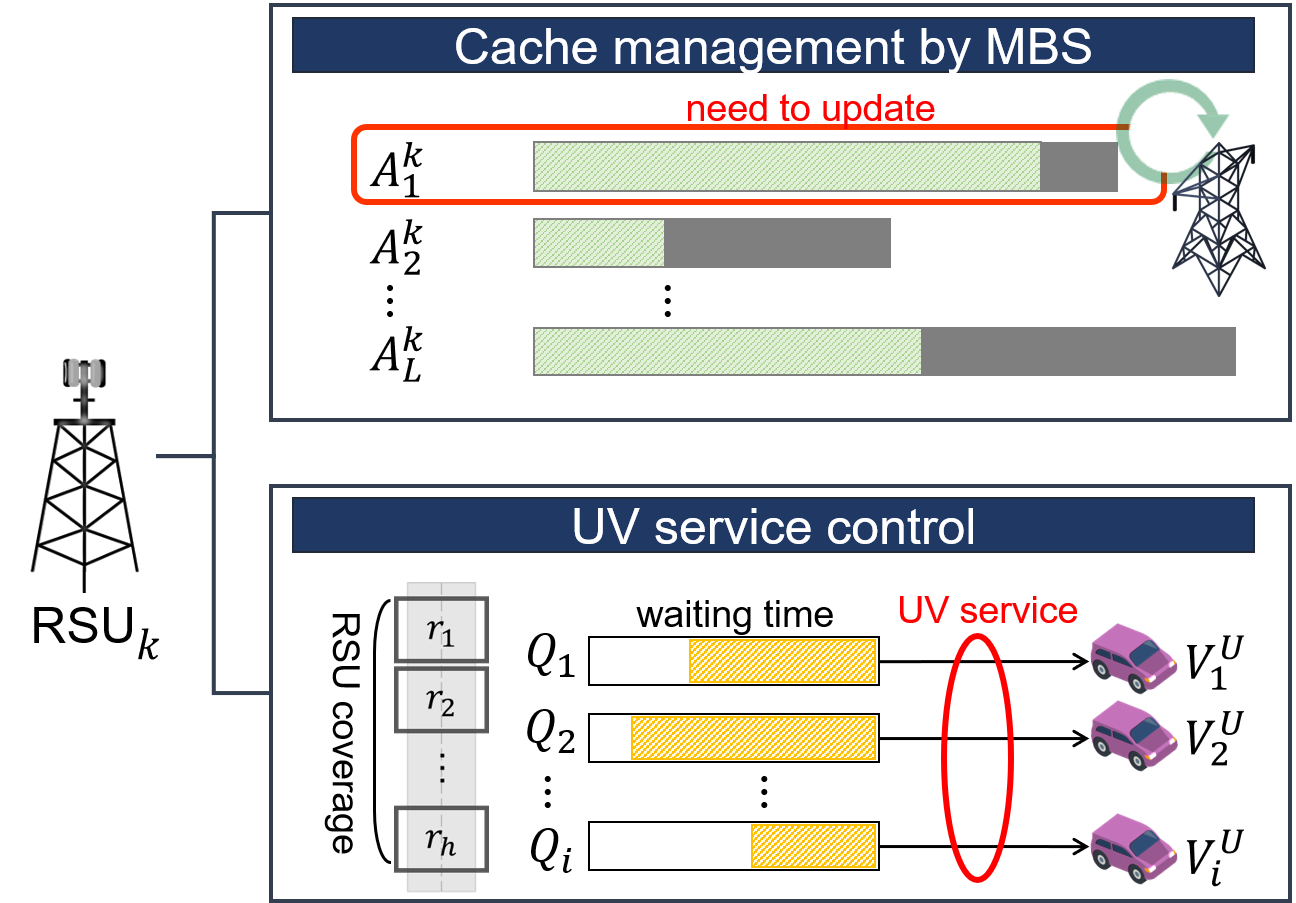}
    \end{center}
    \caption{Role of MBS and RSUs in connected vehicle networks.}
    \label{fig:RSU}
\end{figure}

Suppose that $N_u$ UVs, $N_c$ CVs, $N_R$ RSUs, and one MBS are deployed around the straight road with $L$ regions. Fig.~\ref{network architecture} represents an overall network architecture. The MBS exists in the center of the network and several RSUs are listed along the road at regular intervals. 
% UV와 CV에 대한 설명
The UVs and CVs move in one direction, and the state of the road through which each vehicle (UV or CV) passes is different for each region, such as traffic jam or accident occurrence. 
The UVs request the RSU for the content of a target region within the section of the RSU to which it belongs. 
The target region is an area that the UV wants to check before passing through by receiving content while driving. The position and need of each UV is different, the tolerance time for the request is all different based on the environment. 
The CVs produce road environment data path through all of the road region repeatedly and the produced contents are matched to the region past such as Fig.~\ref{fig:CV}. In CVs' storage, the old content which is not sent to MBS until the content AoI reaches to the maximum limit $A^{max}_h$ is thrown away. 
% CV가 만드는 content에 대한 설명
We assume that all of the contents in CVs are the same size and quality. Depending on the complexity of the road condition the maximum effective time of the content for the region $A^{max}_h$ only varies.
$A^{max}_h$ is a value which is able to be a criterion. This means that the old degree of the content $h$ which contains road information of the region $h$ is valid up to $A^{max}_h$.
% MBS에 대한 설명
In the system, the transmitted contents of CVs are saved in the MBS and delivered to the RSUs and UVs such as Fig.~\ref{fig:MBS} and Fig.~\ref{fig:RSU}. In detail, the MBS receives road content from CVs passing through the road regions and stores content for the road environment. 
%Fig.~\ref{MBS} shows the relationship between CV and MBS. 
The MBS manages the cached contents of RSUs considering the freshness of the distributed contents in the RSUs. If there is a content which AoI value is similar to $A^{max}_h$ in the MBS storage, it must be changed to more fresh one before the timeworn content could be delivered to UVs through RSUs.
% RSU에 대한 설명
%As shown in Fig.~\ref{RSU}, 
The role of RSUs is a distributed cache and UVs service provider. Each RSU caches some contents and receives UV content requests only in the coverage of itself. 
The popularity of the contents of each RSU varies from time to time. There is a limit to the number of connectable channels with UVs. For these reasons, the decision on which UV request should be handled has significant also implications within the entire system based on the cache status managed by MBS.

\subsubsection{Content AoI}

The content freshness of road environment information which is present in the system is represented as AoI value such as $A^c_{j,h}(t), A_h(t)$ and $A^R_{k,h}(t)$ for the road region $r_h$. 
Each value refers to the AoI value for the same content of region $r_h$ in j-th CV ($V^{c}_j$), MBS, and k-th RSU ($R_k$) in order. 
The AoI of the RSUs ($A^{R}_{k,h}$) and the AoI of the MBS ($A_{h}$) are influenced by AoI of content produced in CVs ($A^{c}_{j,h}$) and the value continues to increase over time $t$ if there is no update or upload as a new version for the same regions. We define the AoI values as follows:
\begin{eqnarray}
    & &
    A^c_{j,h}(t+1) = \left \{\begin{array}{c} A^c_{j,h}(t) + 1 \\ 0 \end{array} \right., \forall h \in L, \forall j \in N_c\\
    & &
    A_h(t+1) = \left \{\begin{array}{c} A_h(t) + 1 \\ A^c_{j,h}(t) \end{array} \right., \forall h \in L\\
    & &
    A^R_{k,h}(t+1) = \left \{\begin{array}{c} A^R_{k,h}(t) + 1 \\ A_h(t) \end{array} \right., \forall h \in L, \forall k \in N_R \label{eq:AoI formulas}\\
    & &
    A_h^j, A_h, A_h^k \in \{0, 1, 2, ...,A^{max}_h\}  \forall h \in L
\end{eqnarray}

As mentioned above, all of the regions have different states and different maximum AoI capacities named as $A_h^{max}$. $A^c_{j,h}(t)$ is valid after the content is produced by $V^c_j$. 
The value has $1$ at the first time and increases by the size of the time slot over time. 
When $A^c_{j,h}(t)$ equals to $A_h^{max}$ and the content $C^c_{j,h}$ is not uploaded to the MBS, the CV $V^c_j$ deletes the content from its storage and $A^c_{j,h}(t)$ resets as $0$. 
$A_h(t)$, AoI of the content of region $h$ stored in the MBS, is defined by a decision of content uploading from CVs. If a content for the same region $r_h$ is uploaded from any CVs to the MBS, $A_h(t+1)$ is replaced as $A^c_{j,h}(t)$ due to the content $C^c_{j,h}$ which is produced for the $h$-th region by $j$-th CV is transmitted perfectly after one time slot. Otherwise, $A^c_{j,h}(t)$ increases by $1$. 
$A^R_{j,h}(t+1)$ is impacted to $A_h(t)$. Similar to AoI of MBS, the value is replaced only when content update occurs by the MBS which provides road environment content to RSU. 
If not, the value increases linearly with the flow of time steps.
Unlike $C^c_{j,h}$, $C^{R}_{k,h}$ and $C_h$ which are stored in RSUs and MBS are not thrown away even if each AoI is beyond the maximum value $A^{max}_h$. If $A_h(t)$ and $A^R_{j,h}(t)$ values are larger than $A^{max}_h$, that  means just MBS or RSUs continue to have the content that does not reflect the latest road environment conditions that has passed a long time since it is produced from CV.

\subsection{Problem Formulation}\label{sec:system model-problem}

% overall object에 대한 optimization 수식
% part 1,2를 모두 포함하는 연구 목적을 수식으로 제시
For the fresh content providing in connected vehicle network, we set $4$ values which have to be considered. In~\eqref{eq:utility}, each value means (i) AoI of contents that exist in all RSUs, (ii) communication cost used by the MBS for RSU cache management (content upload from CVs and content update to RSUs), (iii)service waiting delay of UVs, and (iv) communication cost that RSU uses while providing UV service. They are divided as two content transmission stages as mentioned in Sec.~\ref{sec:introduction}. 

\begin{equation}
    \mathcal{V}= \underbrace{{A}^{RSU}(t) + {C}^{MBS}(t)}_{\text{for content caching}} + \underbrace{{D}^{UV}(t) + {C}^{RSU}(t)}_{\text{for content service}}
    \label{eq:utility}
\end{equation}

The overall object which has to be achieved to ensure the latest status of content and prevent indiscriminate communication in the proposed network is as shown in following~\eqref{eq:overall object}.
\begin{eqnarray}
    \min: & & \lim_{T\rightarrow\infty}
    \frac{1}{T}\sum_{t=1}^{T}\mathcal{V}.
    \label{eq:overall object}
\end{eqnarray}

 In the Internet-of-Everything (IoE) era, device-to-device (D2D) communications has important roles in multiple scenarios, when the 5G networking infrastructure has been destroyed or is unavailable. These situations are referred to as infrastructureless D2D (iD2D) communications, where the iD2D mobile equipments (iMEs) establish, maintain, and manage their connections themselves. Since no coordinator provides support in these situations, security controls experience serious problems in terms of authentication, authorization, and privacy. In this paper, we adapt a prefetched asymmetric authentication (pFAA) mechanism as a countermeasure against these challenges. Security analysis proves that the pFAA mechanism protects itself against recent adversary models in the literature.
\begin{figure}[h]
    \begin{center}
        \includegraphics[width=0.95\linewidth]{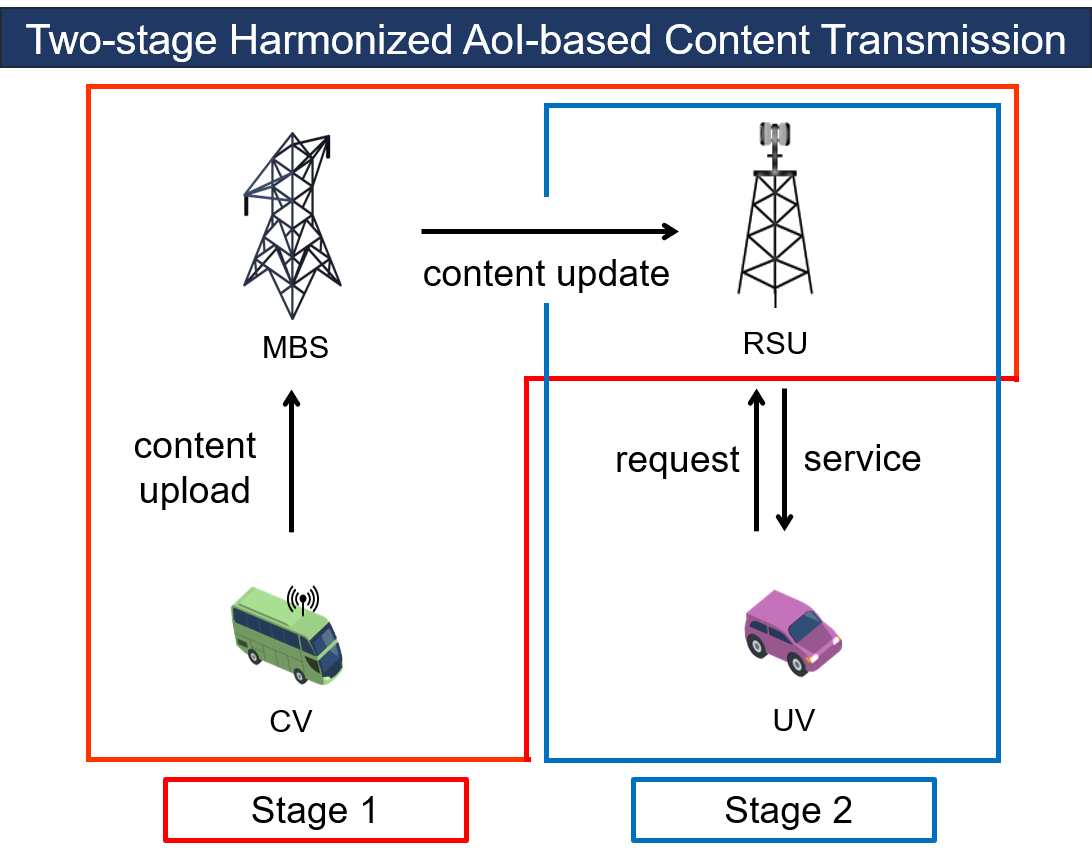}
    \end{center}
    \caption{Two-stage content transmission. The first stage for content caching contains (i) content upload and (ii) content update among CVs, MBS, and RSUs. The second stage for content service contains (iii) content transmission for the response to UV's request.}
    \label{fig:two-stage content transmission}
\end{figure} 
 
To complete~\eqref{eq:overall object}, we suggest a new content caching and service mechanism such as Fig.~\ref{fig:two-stage content transmission}, two-stage harmonized AoI-based contents transmission decision.
The problem is solved and optimized by two independent suggestion algorithms.
The first stage is about optimal content caching through the relationship between CVs, MBS, and RSUs.
The second stage is delay aware content request achievement optimization between RSUs and UVs by guaranteeing the serviced contents' validity. 
The two algorithms are not joint optimization and independent of each other, but not completely separate ideas.\\
As we can see in content AoI formulas~\eqref{eq:AoI formulas},  the updated content AoI of RSU is influenced by MBS and CVs. 
The content state of the RSU, which is determined according to the result of the first stage, is transferred to the UVs in the second stage.
For this reason, content caching and service are all important for the proposed system.
The algorithms guarantee caching and serving optimality in the network environment proposed for the goal of caching for content service considering AoI, respectively. 
From the next section, the algorithms for the two stages are described sequentially.

\section{Optimal Upload and Update for the Freshness of Cached Content}\label{sec:proposed1}
The MBS at the center of the network receives a newly generated road contents from the CVs and updates the old contents of the RSUs.
In this section, we propose an algorithm that determines which contents of CVs will be uploaded and which contents of RSUs will be updated for the RSU cache management.

\subsection{AoI-Aware Contents Caching}
The object of the first stage which is optimal content caching is able to be formulated as follows: 
\begin{eqnarray}
    \min_{x, y}: & & \lim_{T\rightarrow\infty}
    \frac{1}{T}\sum_{t=1}^{T}\left(A^{RSU}(t) + C^{MBS}(t)\right)
    \label{eq:MDP object}
\end{eqnarray}

This problem formulation to minimize content AoI of all RSUs and communication cost of MBS for uploading and updating~\eqref{eq:MDP object} is separated from~\eqref{eq:utility}. In the formulation, the smaller the two values, the more satisfying the purpose. 
However, in the system that we assume, for the $A^{RSU}(t)$, the ratio that means how much scope is to the max value $A^{max}_h$ is more important than the absolute number size. For this reason,~\eqref{eq:MDP object} can be transformed as a problem that maximize the values defined as an utility. The utility is a combination of the current AoI value compared to $A^{max}_h$ and the communication cost at time stet $t$. The modified equation is as follows:

\begin{eqnarray}
    \max_{x, y}: & & \sum_{t=1}^{T}\mathcal{U}(t)\\
    \text{s.t.} & &    
    \mathcal{U}(t) = \epsilon\cdot(\mathcal{U}^{RSU}_{AoI}(t)w) - (1-\epsilon)\cdot\mathcal{U}^{MBS}_{cost}(t) \label{eq:stage1 utility}\\
    & &
    \mathcal{U}^{RSU}_{AoI}(t) = \sum^{N_R}_{k=1}\sum^{L}_{h=1} \frac{A^{max}_h}{A^R_{k,h}(t)}\cdot W\cdot p^k_h(t)  \label{eq:stage1 AoI utility}\\
    & &
    \mathcal{U}^{MBS}_{cost}(t) = \sum^{N_c}_{j=1}\sum^{L}_{h=1}C^j_h(t) + \sum^{N_R}_{k=1}\sum^{L}_{h=1}C^k_h(t)  \label{eq:stage1 cost utility}\\
    & &
    % (A_h(t-1) - A^R_{k,h}(t-1))y^k_h(t) \leq \nonumber \\
    % & & \quad\quad\quad\quad\quad A^{max}_h-(A^R_{k,h}(t-1) +1), \forall h, k, t \label{eq:stage1 AoI (i)}\\
    % & &
    % \sum^{N_C}_{j=1}(A^j_h(t-1) - A_h(t-1))x^j_h(t) \leq \nonumber \\
    % & & \quad\quad\quad\quad\quad A^{max}_h-(\sum^{N_C}_{j=1}A_h(t-1)+1), \forall h, j, t \label{eq:stage1 AoI (ii)}\\
    % & &
    A^R_{k,h}(t) = (1-y^k_h(t))\cdot(A^R_{k,h} (t-1) + 1) \nonumber\\
    & & \quad\quad\quad\quad\quad + y^k_h(t)\cdot A_h(t-1), \forall k, h\in N_R, L \label{eq:stage1 AoI (i)}\\
    & & 
    A_h(t) = \sum_{\forall j\in N_C}\{(1-x^j_h(t))\cdot(A_h(t-1)+1) \nonumber\\
    & & \quad\quad\quad\quad\quad + x^j_h(t)\cdot A^c_{j,h}(t-1)\}, \forall h\in L \label{eq:stage1 AoI (ii)}\\
    & &
    C^c_{j,h}(t) = x^j_h(t)\cdot d_j(t), \forall j, h\in N_C, L \label{eq:stage1 cost-cv} \\
    & &
    C^R_{k,h}(t) = y^k_h(t)\cdot d_k(t) \cdot \frac{1}{p^k_h(t)}, \forall k, h\in N_R, L \label{eq:stage1 cost-rsu}\\
    & & 
    \sum^{N_c}_{j=1}x^j_h(t) \leq 1, \forall h\in L\\
    & & 
    \sum^{L}_{h=1}x^j_h(t) \leq 1, \forall j\in N_c\\
    & & 
    \sum^{L}_{h=1}y^k_h(t) \leq 1, \forall k\in N_R\\
    & &  
    \sum_{\forall j\in N_C}\sum_{\forall h\in L}x^j_h(t) + 
    \sum_{\forall k\in N_R}\sum_{\forall h\in L}y^k_h(t) \leq H\\
    & &
    x^j_h(t), y^k_h(t) \in [0,1], \forall j, k, h
\end{eqnarray}
As mentioned above, the main object function of the first stage is replaced to maximize the utility which is decided by the utilities of content AoI value and communication cost. By the constrains~\eqref{eq:stage1 AoI utility},~\eqref{eq:stage1 cost utility}, the smaller ${A}^{RSU}$ and ${C}^{MBS}$, the greater each utility. In~\eqref{eq:stage1 utility}, $\epsilon$ is an importance ratio for content age in RSUs and transmission cost of MBS. $w$ is a value to match the size of two values.
The measurement of content AoI utility is judged to be the comparison between the maximum value that can recognize the validity of the data and the current value by \eqref{eq:stage1 AoI utility}. In addition, $W$ means the weight value of the $h$-th content of RSU $k$ compared to all of the content AoI values in the system at time step $t$. 
The communication cost utility is the sum of costs occurring in two cases, content uploading and updating.
In~\eqref{eq:stage1 AoI (i)}--\eqref{eq:stage1 cost-rsu}, each value is determined by two variables (i.e., $x^j_h(t), y^k_h(t)$). 
%For the AoI values of RSUs and MBS,~\eqref{eq:stage1 AoI (i)} and~\eqref{eq:stage1 AoI (ii)} ensure tat the AoI values after the upload and update are not over the maximum values of the contents.
\eqref{eq:stage1 cost-cv} and~\eqref{eq:stage1 cost-rsu} determine the communication costs for the content transmission of MBS and RSUs. Since, we assume the content file size is equal to all of the regions, the cost is determined by the distance to the selected target and the bandwidth size used. The popularity of content is reflected, especially for RSUs that directly support UVs.
Each variable means whether to transmit content between CVs and MBS, and between MBS and RSUs. $x^j_h(t)$, the decision variable for uploading, has conditions that only one content can be uploaded in one CV, and that multiple CVs does not duplicately upload for the same content.
$y^k_h(t)$, the decision variable for updating, is limited to the condition that only one content can be updated in one RSU.
In addition, we restrict the number of CVs and RSUs that MBS can connect at the same time to the channel limit $H$.

% \begin{figure}[ht]
%     \begin{center}
%         \includegraphics[width=0.75\linewidth]{image/MBS_decision_control.png}
%     \end{center}
%     \caption{Content transmission decision of MBS. 
%     MBS has two kinds of control actions ($x^j_h$ and $y^k_h$). For the content upload ($x^j_h$), only one content is transmitted from one CV without overlapping for the same content. The content update for each RSU ($y^k_h$) is occurs only for one content in each RSU.}
%     \label{fig:MBS_decision_control}
% \end{figure}

\subsection{Formulation with Markov Decision Process (MDP)}
To solve the above optimization problem, we utilize an MDP model $<\mathcal{S}, \mathcal{A}, \mathcal{P}, \mathcal{R}, \gamma>$ which guarantees the optimal solution for every moment. 
Therefore, in this part, we characterize the dynamic vehicle content caching network environment as follows:

\BfPara{State Space} Information used by MBS, an agent, in an environment to which MDP is applied, is described. The state contains AoI of all contents in the system, distance between system components and agents, channel state of itself, and the contents' population that each RSU has
\begin{equation}
    \mathcal{S}(t)=\{[A(t)], [d(t)], [h(t)], [p(t)]\}
\end{equation} 
where
\begin{itemize}
    \item $[A(t)]$ consists of $A^c_{j,h}(t)$, $A_h(t)$, $A^R_{k,h}(t)$, and $A^{max}$ where AoI values for content $h$ stored in CV $j$, MBS and RSU $k$ depending on the action $x$ and $y$. Lastly, $A^{max}$ is a maximum AoI value equally assigned to CV $j$, MBS, and RSU $k$.
    \item $[d(t)]$ consists of $d_j(t)$ and $d_k(t)$ where the distance from MBS to CV $j$ and the distance from MBS to RSU $k$, respectively.
    \item $[h(t)]$ stands for the channel state of MBS which is determined by action $x$ and $y$.
    \item $[p(t)]$ stands for the popularity of RSU $k$'s content $h$.
\end{itemize}

\BfPara{Action Space} In this MDP environment, actions replace the two variables, $x^j_h(t)$ and $y^k_h(t)$. The two actions are binary variables and each meaning is as follows,
\begin{equation}
    \mathcal{A}(t) = \{[x(t)], [y(t)]\}
\end{equation}
where
\begin{itemize}
    \item $[x(t)]$ consists of $x^j_h(t)$ which are binary indices whether the content $h$ is uploaded or not from CV$_j$ to MBS.
    \item $[y(t)]$ consists of $y^k_h(t)$ which are binary indices whether the content $h$ in RSU $k$ is updated or not by MBS.
\end{itemize}

\BfPara{Transition Probability} The transition probability function is formulated as following~\eqref{eq:trans probabiltiy} where the function means that the agent will be convert to the next state $s(t+1)$ when taking an action $a(t)$ from the current state $s(t)$ with the probability of~\eqref{eq:trans probabiltiy}.
\begin{equation}
    P(s(t+1)\mid s(t), a(t)) \label{eq:trans probabiltiy}
\end{equation}

\BfPara{Reward Function} The reward function is equal to the first constraint of the optimization formula~\eqref{eq:stage1 utility}. 
Reward function is set to maximize utility of content caching so that the agent MBS determines the appropriate actions, i.e., $x^j_h(t)$ and $y^k_h(t)$,
\begin{equation}
\begin{split}
    r(s(t), a(t)) &= \mathcal{U}(t) \\ &=\epsilon\cdot (\mathcal{U}^{RSU}_{AoI}(t)\cdot w) - (1-\epsilon)\cdot \mathcal{U}^{MBS}_{cost}(t)
\end{split}
\end{equation}
where $\epsilon$ stands for the weight between two factors, i.e., content AoI and communication cost. In this paper, we assume the two factors are equally considered, i.e., $\epsilon=0.5$.
More details about $\mathcal{U}^{RSU}_{AoI}$ (utility for content AoI) and $\mathcal{U}^{MBS}_{cost}(t)$ (utility for communication cost) are as follows.

\begin{itemize}
    \item $\mathcal{U}^{RSU}_{AoI}(t)$: Equivalent to~\eqref{eq:stage1 AoI utility},
it stands for the proportion of the current AoI value of the RSU to the reference value $A^{max}_h$. In~\eqref{eq:stage1 AoI utility}, $A^R_{k,h}(t)$ is affected by the two actions (i.e, $x^j_h(t)$ and $y^k_h(t)$ according to~\eqref{eq:stage1 AoI (i)} and~\eqref{eq:stage1 AoI (ii)}.
If the maximum AoI of two contents are different (e.g.,$A^{max}_1 = 7$ and $A^{max}_2 = 4$ ) and the contents have the same AoI value at time step t (e.g.,$A^{R}_{k,1}(t) = A^{R}_{k,2} = 3$), the utility for the first content is greater than the second content utility.
% \begin{multline}
%     A^R_{k,h}(t) = (1-y^k_h(t))\cdot(A^R_{k,h} (t-1) + 1) \\+ y^k_h(t)\cdot A_h(t-1), \forall k, h\in N_R, L
%     \label{eq:AoI_RSU}
% \end{multline}
% \begin{multline}
%     A_h(t) = \sum_{\forall j\in N_C}\{(1-x^j_h(t))\cdot(A_h(t-1)+1) \\+ x^j_h(t)\cdot A^c_{j,h}(t-1)\}, \forall h\in L
%     \label{eq:AoI_MBS}
% \end{multline}

    \item $\mathcal{U}^{MBS}_{cost}(t)$: Equivalent to~\eqref{eq:stage1 cost utility},
and it is also affected by the actions and accumulated only when each action value equals to $1$. In~\eqref{eq:stage1 cost-rsu}, we apply the content popularity at time step $t$ as $p^k_h(t)$. Its intention is to ensure that even if frequent communication occurs for the freshness of content, if the content is popular from UVs that the RSU should service, $C^R_{k,h}(t)$ has a smaller value than other cases.
\end{itemize}
\BfPara{Value Function} The object of the MDP-based content caching is to achieve optimal content transmission decisions between the AoI of the contents present in the system and the communication cost according to the content movement. We define $\pi:S\rightarrow A$ which maps the current state with series of actions,(e.g., $a = \pi(s)$). We denote $\Pi$ is a set of all stationary policies. For any initial state $s$ and corresponding policy $\pi \in \Pi$, the cumulative reward during $T$ time-step is formulated as follows:
\begin{equation}
    \max_{\pi \in \Pi}:\sum_{t=1}^{T}\gamma^{t}r(s^\pi(t), a(t))\\
\end{equation}
where the discount factor $\gamma$ has a value in $[0,1]$. Based on the transition probability and cumulative reward, the value function $V$ is defined as 
\begin{equation}
    V^*(s) = \max_{a\in A}\{r(s,a) + \gamma^t\sum_{s'\in S}P(s'\mid s,a)V^*(s')\} \label{eq:value function}
\end{equation}
where $s$ and $a$ are the current state and caching action at the time slot $t$, and $s'$ is the next state by that action at the time slot $t+1$. The Bellman equation~\eqref{eq:value function} is solved using traditional value or policy iteration and the process is presented in Algorithm~\ref{algo:MDP_caching}.

\begin{algorithm}[t!]
\small
%\setstretch{1.5}
\caption{AoI aware content caching}
\label{algo:MDP_caching}
 \textbf{Input:} reward function $r(s(t), a(t))$, transitional model $P(s'|s,a)$, discounted factor $\gamma$, convergence threshold $\theta$\\
 \textbf{Output:}optimal policy $\pi^{*}$\\
 Initialize $V(s)$ with zeros\\
 Converge $\leftarrow$ false\\
 \While{converge $=$ false}{
    $\Delta \leftarrow 0$ \\
    \For{$s \in S$}{
        temp $\leftarrow v(s)$
        $v(s) \leftarrow r(s,a) + \gamma^t\sum_{s'\in S}P(s'\mid s,a)V^*(s')$
        $\Delta \leftarrow \max(\Delta, |temp-v(s)|)$
        }
    \If{$\Delta <\theta$}{
        converge $\leftarrow$ true
        }
    }
 \For{$s\in S$}{
    $\pi^{*}(s) \leftarrow argmax\sum_{s'\in S}P(s'\mid s,a)V^*(s')$
    }
 \textbf{Return} $\pi^{*}$
\end{algorithm} 

\section{Content Request Achievement Optimization}\label{sec:proposed2}
We assume the situation that several UVs request contents for particular region to RSU as it passes through the region of the road and RSU determines whether to send the content (service) at current time for multiple requests it receives. 

\subsection{Lyapunov Optimization}
For the content service in RSU, the content AoI served to the UVs and the communication cost between the RSU and UVs are considered. 
Since UVs request for a specific content as needed, unlike section~\ref{sec:proposed1}, new constraints are added for RSU to quickly support UVs. Depending on the location of the UV or the requested content, there is an importance of preventing excessive latency for each UV, and the content which is transmitted after a specific allowable delay dose not become valid data for the UV driving on the road.
Therefore, the condition for the waiting time from the occurrence of the UV request to the service is considered. 
For this reason, we present a Lyapunov optimization-based RSU control algorithm to meet all three considerations: Content AoI with RSU's communication cost and UV's delay.
As mentioned above, we deal with content AoI and communication cost for the content service in~\eqref{eq:utility} through Lyapunov optimization.
\begin{eqnarray}
    \min: & & \lim_{T\rightarrow\infty}
    \frac{1}{T}\sum_{t=1}^{T}(D^{UV}(t) + C^{RSU}(t))
    \label{LO object}
\end{eqnarray}

To satisfy the goal of~\eqref{LO object}, we replace \eqref{LO object} by

\begin{eqnarray}
    \min: & & \lim_{T\rightarrow\infty}\sum_{t=1}^{T}\sum_{i=1}^{N'_U} C_i(\alpha_i[t])\label{eq:opt} \\
    \text{s.t.}
    & &
    \lim_{T\rightarrow\infty}\sum_{t=1}^{T}\sum_{h=1}^{L'}D_i[t]\cdot r_{i,h}[t]<\infty, \forall i \in N'_U\\
    & &
    \sum_{h=1}^{L'}A^{rx}_i(\alpha_i[t])\leq A^{max}_hr_{i,h}[t], \forall i \in N'_U \label{eq:Lyapunov_AoI_const}\\
    & &
    C_i(\alpha_i[t]) = \sum_{h=1}^{L'}r_{i,h}[t]\cdot \alpha_i[t]\cdot d_i[t], \forall i \in N'_U \label{eq:Lyapunov_cost}\\
    & &
    A^{rx}_i(\alpha_i[t]) = (\sum_{h=1}^{L'}r_{i,h}[t]A'_h[t] + 1)\cdot \alpha_i[t], \forall i \in N'_U \label{eq:Lyapunov_AoI}\\
    & &
    \alpha_i[t], r_{i,h}[t] \in \{0, 1\}\\
    & &
    \sum^{L'}_{h=1}r_{i,h}[t] = 1, \forall i \in N'_U\\
    & &
    \sum^{N'_U}_{i=1}\alpha_i[t] \leq H^{UV}
\end{eqnarray}
\begin{figure}[t]
    \begin{center}
        \includegraphics[width=0.95\linewidth]{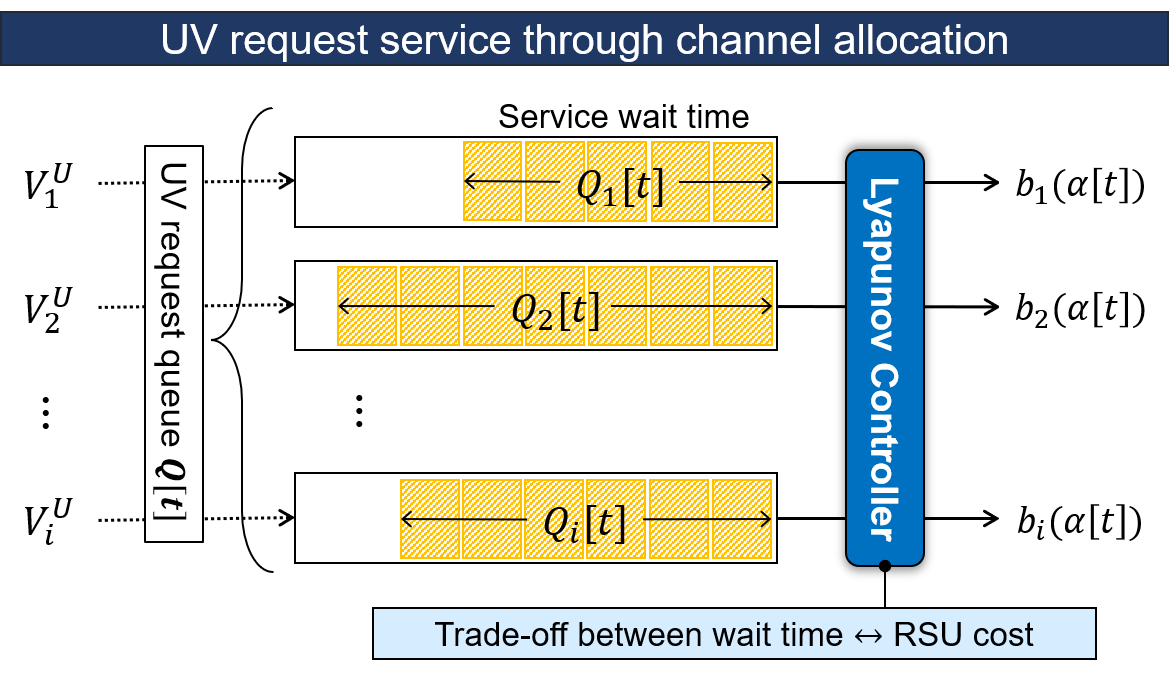}
    \end{center}
    \caption{Lyapunov optimization-based UV service control.}
    \label{fig:RSU_laypunov}
\end{figure}  

When there are $N'_U$ UVs in one RSU coverage, a queue for the UV which requests a specific content to the RSU is set as waiting queue such as Fig.~\ref{fig:RSU_laypunov}. Each RSU has several queueus for the UVs which request contents to the RSU. The waiting queue exists only when UV sends a request. After UV sends the request, the waiting time is accumulated in the queue until service is completed.
The RUS decides for all the waiting queues whether to support each UV so that the matched waiting queue does not overflow.
In the proposed Lyapunov optimization, it indicates that the RSU does not consider only one UV, but also comprehensive control over all UVs present in the coverage of the RSU. 
In the process of serving UV, we focus heavily on transmitting content that has not expired within an acceptable time of UVs, rather than ensuring that it always delivers the latest content.
The waiting queue and AoI utility are valid only when the UV requests content $(r_{i,h} = 1)$. 

The purpose of the expression is to minimize the communication cost required for RSU to service UVs while satisfying delay ($D_i[t]$) and AoI ($A^{max}_hr_{i,h}[t]$) constraints. It prevents overflow of the waiting time queues for each UV which sends the content request and ensures that the AoI of the transmitted content does not exceed the maximum value (i.e., the content remains valid for UV use when the requested content is sent to UV).
In the formulations, $r_{i,h}[t]$ means whether UV$_i$ has requested content $h$ to the RSU in time step $t$, and $\alpha_i[t]$, which means whether the RSU will serve that UV, is valid only if the value of $r_{i,h}[t]$ is $1$. 
$C_i(\alpha_i[t])$ and $A^{rx}_i(\alpha_i[t])$ are determined by the control action of RSU $\alpha_i[t]$ for $N'_U$ UVs and are the values to be adjusted initially intended as shown in~\eqref{eq:utility} through this study.

In~\eqref{eq:Lyapunov_cost}, $C_i(\alpha_i[t])$ is calculated according to the distance between UV$_i$ and RSU and the fixed bandwidth size used, similar to the communication cost between CV and MBS. The AoI of the content which UV receive is calculated by adding $1$ time slot required for content transmission to the AoI value of the cached data in the RSU at the time point as shown in~\eqref{eq:Lyapunov_AoI}. 
In the system, we assume that only one content can be requested when each UV sends a single request to the RSU and there is limit to the number of UVs that can be supported simultaneously by one RSU. 
Through the above conditions, the RSU should derive the optimal action for each UV that can minimize its own communication cost, taking into account the waiting queue and received content AoI using $H^{UV}$ constrained channels.

\subsection{Queue/Delay-based Optimal Control}
\begin{figure}[t]
    \begin{center}
        \includegraphics[width=0.75\linewidth]{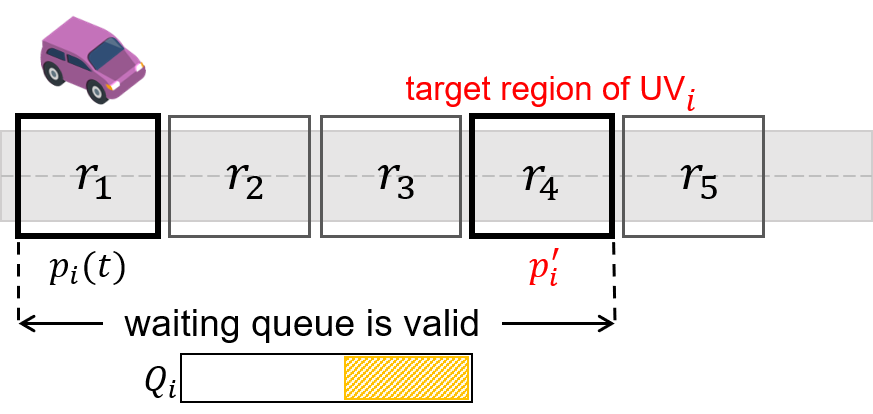}
    \end{center}
    \caption{Validity of the queue according to the UV location.}
    \label{fig:waiting queue}
\end{figure}
In this section, we present a queues for service waiting time (i.e., delay ($D_i[t]$) ) already mentioned above. 
Modeling for the waiting queue is as follows:

\begin{multline}
    Q_i[t+1] = \max\big[Q_i[t]-b(\alpha_i[t]), 0\big] + a[t] \\, p'_i-p_i[t] \geq 0
    \label{eq:queue validity }
\end{multline}

In each RSU, there are several waiting queues for UVs which request content path through the road shch as Fig.~\ref{fig:waiting queue}. The value of $p_i[t]$ means the region number where the $UV_i$ belongs and $p'_i$ means the target region number which is requested to get road content by the $UV_i$. It has the same set as the number of the region $(p_i \in {1, 2, \cdots, L})$.
$Q_i[t]$ is the waiting delay of $UV_i$ which is generated after the UV request some content to the RSU. One RSU has a maximum $N'_U$ multi-waiting queues for the request of the UVs in the coverage at the same time. 
The queue is about the delay expressed in time, the arrival and departure is calculated by time.
Since the queue has been created, the waiting time is automatically accumulated as arrival process of the queue. The departure of queue is affected by the decisions of RSU that represents whether to provide services to each UV through control action. 
The waiting delay of $UV_i$ modeled as queue is meaningful only when the UV is in the same region or the previous region $(p'_i-p_i[t] \geq 0)$ as shown in Fig.~\ref{fig:waiting queue}. If the UV passed the region before it receives the request service, the waiting queue has no reason to exist.
The waiting queue of the UV that has passed the valid region is automatically removed from the RSU and the process is applied until the UV in the coverage of the RSU leaves the boundary. When UV sends a new request, the queue becomes valid again.
%$Q^i_{AoI}[t]$ is the virtual AoI queue about the content requested by the $UV_i$. It is different from the delay queue where time is accumulated according to the time step, the virtual queue has no arrival process and only has departure process which represents the maximum capacity of the content freshness limitation. By setting the virtual queue as above, it guarantees that the transmitted content is a valid road information when the content forwarded to the UV has an AoI value less than $A^{max}_h$. $\alpha_i[t]$ is the channel allocation control action for the $UV_i$ which sends the road information request. Based on the waiting queue and the virtual AoI queue status, the controller decides whether the service is supported at the current time for the $UV_i$ by achieving \eqref{eq:opt}. 

In \eqref{eq:opt}, $C_i(\alpha_i[t])$ stands for the RSU communication cost for each UV delay queue departure process $b_i(\alpha_i[t])$ when the given channel allocation decision is $\alpha_i[t]$. As mentioned earlier, the channel allocation for the content service decision generates a trade-off between the minimization of communication cost and stability of the queuing system which is related to the average delay (i.e., each UV's service waiting time). 

Respect to this trade-off, the Lyapunov optimization theory-based drift-plus-penalty (DPP) algorithm~\cite{mm2017jonghoe,tmc2019jonghoe} is applied for optimizing the time-average utility function (i.e., communication dost) subject to queue stability.
Define the Lyapunov function 
$L(Q[t]) = \frac{1}{2}\sum_{i=1}^{N^U_k}(Q[t])^2$,
and let $\Delta(.)$ be a conditional quadratic Lyapunov function that can be formulated as 
\begin{eqnarray}
    \mathbb{E}[L(Q_i[t+1])-L(Q_i[t])| Q_i[t]]
\end{eqnarray}

called as the drift on $t$. 
After the MBS drone $i$ where $\forall i \in\mathcal{M}$ observes the current queue length $Q_i(t)$, the channel to support content transmission is required in each time slot.
According to \cite{book2010sno}, this dynamic policy is designed to achieve queue stability by minimizing an upper bound on drift-plus-penalty which is given by
\begin{equation}
    \Delta(Q_i[t]) + V \mathbb{E} \Big[ C_i(\alpha_i[t]) \Big],
\end{equation}
where $V$ is an importance weight for communication cost minimization. 
The following is a process of induction for the upper bound on the drift of the Lyapunov function on $t$:
\begin{multline}
    L(Q_i[t+1]) - L(Q_i[t]) = 
    \frac{1}{2}\Big( Q_i([t+1]^2 - Q_i[t]^2 \Big) \\
    \leq \frac{1}{2} \Big( a_i[t]^2 + b_i(\alpha_i[t])^2 \Big) + \\
     Q_i[t] (a_i[t] - b_i(\alpha_i[t])).
\end{multline}

Therefore, the upper bound on the conditional Lyapunov drift can be obtained as follows:
\begin{multline}
    \Delta(Q_i[t]) = \mathbb{E}[L(Q_i[t+1]) - L(Q_i[t]) | Q_i[t]] \\ 
    \leq C + \mathbb{E}\Big[ Q_i[t](a_i[t] - b_i(\alpha_i[t]) \Big| Q_i[t] \Big],
\end{multline}
where $C$ is a constant which can be obviously expressed as
\begin{eqnarray}
    \frac{1}{2}\mathbb{E}\Big[ a_i[t]^2 + b_i(\alpha_i[t])^2 \Big| Q_i[t] \Big] &\leq& C,
\end{eqnarray}
where this assumes that the arrival and departure process rates are upper bounded.
Given that $C$ is a constant and that the arrival process of $a_i[t]$ is uncontrollable, the reduction of the upper bound on drift-plus-penalty takes the following forms:
\begin{equation}
    V \mathbb{E}\Big[ C_i(\alpha_i[t]) \Big] - \mathbb{E}\Big[ Q_i[t]\cdot b_i(\alpha_i[t]) \Big].
    \label{eq:dpp_exp}
\end{equation}

Here, the idea of opportunistically minimizing the expectations is used; and as a result, \eqref{eq:dpp_exp} can be reduced by an algorithm that observes the current delay state $Q_i[t]$ and determines $\alpha_i[t]$ for each UV$_i$ at every slot $t$.
\begin{figure*}
\begin{equation}
    \alpha^{*}_i[t]\leftarrow
    \arg\min_{\alpha_i[t]\in\mathcal{A}}
    \left[V\cdot C_i(\alpha_i[t]) - Q_i[t]b_i(\alpha_i[t])\right]
    , \forall i \in N'_U.
    \label{eq:lyapunov-final}
\end{equation}
\hrulefill
\end{figure*}

In order to verity whether \eqref{eq:lyapunov-final} works as desired, simply two possible cases can be considered as follows, i.e., $Q_i[t]=0$ and $Q_i[t]\approx \infty$.
\begin{itemize}
    \item Suppose that $Q_i[t]=0$. Then, the \eqref{eq:lyapunov-final} tries to minimize $V\cdot C_i(\alpha_i[t])$, i.e., the RSU dose not allocate channel to UV$_i$ in a situation where the algorithm satisfies with condition~\eqref{eq:Lyapunov_AoI_const} due to the waiting time of the UV is not so long there is enough time to wait. This is semantically true because we can focus on the main objective, i.e., communication cost of the RSU, because stability is already achieved at this moment.
    \item On the other hand, suppose that $Q_i[t]\approx \infty$. Then, the \eqref{eq:lyapunov-final} tries to maximize $b_i(\alpha_i[t])$, i.e., RSU allocates a channel to up to $H^{UV}$ UV$_i$ and transmits the requested contents immediately. The accumulated times in the queue are all emptied into the department process $b_i(\alpha_i[t])$ and the queue is possible to maintain stability. This is also true because stability shoud be mainly considered when $Q_i[t]$ even though the RSU use certain amount of communication cost to avoid queue overflow.
\end{itemize}

Finally, we confirm that our proposed closed-form mathematical formulation, i.e., \eqref{eq:lyapunov-final}, controls $\alpha_i[t]$ for minimizing time-average communication cost subject to queue stability. The pseudo-code of the proposed content service optimization algorithm is presented in Algorithm~\ref{alg:Lyapunov}.

\begin{algorithm}[t]
\small
\caption{Queue based optimal service}
\label{alg:Lyapunov}
\textbf{Initialize:} $t\leftarrow 0$, $Q_{i}[t]\leftarrow 0$\\
Decision action: $\forall \alpha_{i}(t)\in{0,1}$ \\
%\textbf{Stabilized Adaptive Ship Detection:}\\
\While{$t\leq T$}{ 
    Observe $Q_{i}(t)$
    $\mathcal{I}^{*} \leftarrow -\infty$
    \For{$\alpha_{i}[t]\in {0,1}$}{
        $\mathcal{I}\leftarrow V\cdot C_{i}(\alpha_{i}(t)) - Q_{i}(t)b_{i}(\alpha_{i}(t))$;
        \If {$\mathcal{I} \geq \mathcal{I}^{*}$}{
            $\mathcal{I}^{*}\leftarrow\mathcal{I}$, $\alpha_{i}^{*}[t]\leftarrow \alpha_{i}[t]$
            }
        }
    }
\end{algorithm}

\section{Performance Evaluation}\label{sec:simulation}
This section describes our simulation setup for performance evaluation and its related evaluation results.

\subsection{Simulation Settings}

\begin{figure}[ht]
    \begin{center}
        \includegraphics[width=0.95\linewidth]{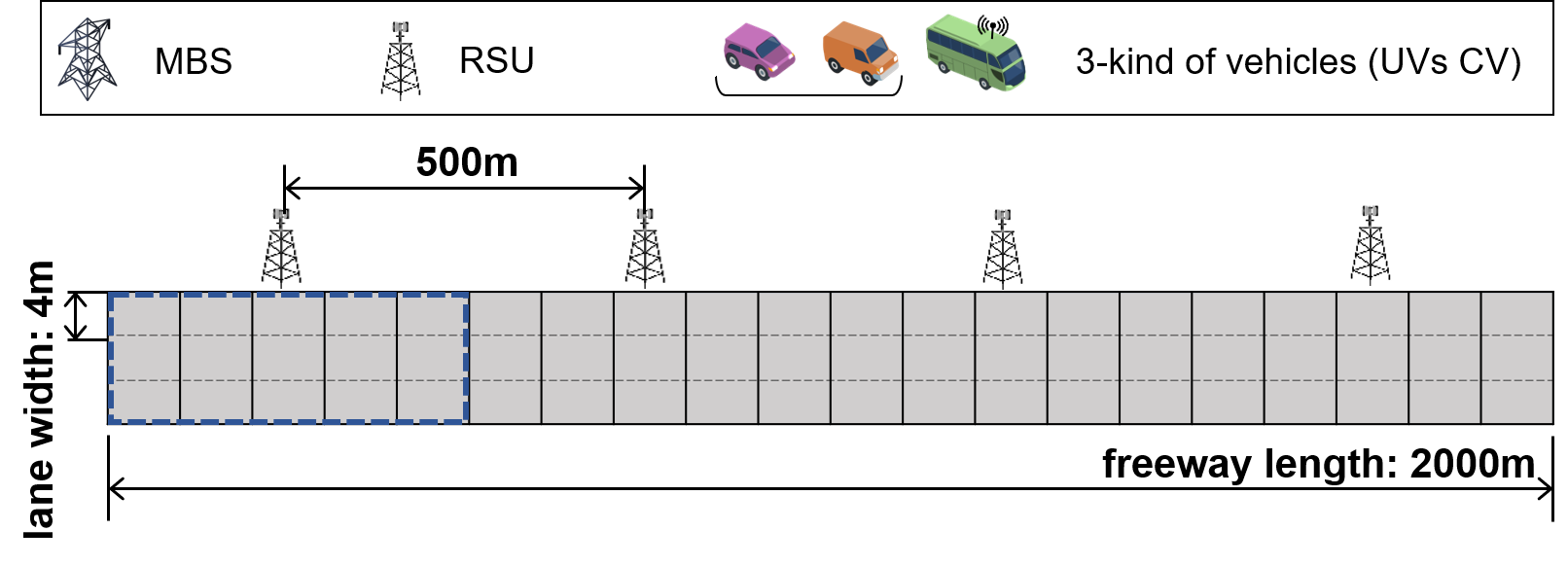}
    \end{center}
    \caption{Road configuration and infrastructure deployment for freeway case.}
    % 3GPP TR 37.885 v15.3.0
    % 3GPP TR 36.885 v14.0.0
    \label{fig:road}
\end{figure}

% 실험 환경 설정
The performance of the proposed two-stage AoI-based content caching and dynamic content service decision method is evaluated by assuming a vehicle network in the road situation as shown in Fig.~\ref{fig:road}.
The high-way environment is constructed with one MBS, $4$ RSUs, and $2$-kind of vehicles (e.g., UV and CV). The road has $3$ lanes, UVs and CVs on the road move at different speeds for each lane.
Whole length of the high-way is $2000$m and each RSU covers $500$m area. The RSU coverage area is divided into $5$ regions and each region is mapping to one content which is cached and managed by the RSU.
We set the initial position of each vehicle randomly and initialize the position when it is out of the road range. The vehicle speed in each lane is basically set as 30, 50, and 80, and the unit is unified as km/h.
Regions through which UV and CV pass have different traffic conditions as previously described, and thus have different content $A^{max}_h$ values.
For the evaluation of the proposed algorithm, the $A^{max}_h$ value for the regions is set to a value less than $20$ (e.g., \{normal: 20, traffic jam: 10, accident: 8, crowded: 15\}). 
The region types are arbitrarily arranged on the road, so that all types of content may not be managed by one RSU at all times.
In addition, the content AoI for each region initially stored or cached in the MBS and RSUs is set to random within the range not exceeding the maximum value $A^{max}_h$.
In the system, for performance evaluation of the proposed MDP-based AoI aware content caching algorithm, we assume that the MBS has totally 6 channel limitation, and it can communication with maximum 3 CVs and 3 RSUs at the same time. 
For the Lyapunov optimization based content service algorithm, we do not specify a separate limit on the number of UVs that the RSU can serve at the same time. Rather than controlling the number of UVs the RSU supports, control whether to provide real-time services for content requests sent by each UV.

% performance metric
As performance metrics in the connected vehicle network, we mainly focus on 1) the updated content AoI (e.g., freshness) for all of the contents in the system, 2) the communication resource usage (e.g., cost) for content caching and service stages, and 3) the queue backlog that measures the service waiting queue stability.
Each element is a concept that is considered importantly in the process of solving~\eqref{eq:overall object} that we describe through Sec.~\ref{sec:proposed1} and Sec.~\ref{sec:proposed2}.

\subsection{Simulation Results and Analysis}

\subsubsection{Performance of the Content Caching Algorithm}

% 비교 알고리즘 2개
% (1) AoI-greedy (AoI-max에 대한 고려 X, 절대적인 AoI 수치만 낮게) --> MBS가 사용가능한 모든 channel을 max로 사용할 것으로 예상
% (2) random 
In this part, we describe the performance results of our proposed AoI aware content caching algorithm compared to the other two algorithms (e.g., AoI-greedy and random algorithm). The random algorithm performs content transmission (uploading and updating) at random; AoI-greedy algorithm performs content caching by considering only lowering the AoI sum of all content in the system without the concept of maximum allowable value $A^{max}_h$. As mentioned above, now we evaluate how up-to-date the content of the RSU is and how much cost is consumed in the process of caching the content.
% ==================== Figure: Average AoI in RSUs ====================

\begin{figure*}[t]\centering
    \begin{multicols}{4}
     \includegraphics[width=0.9\linewidth]{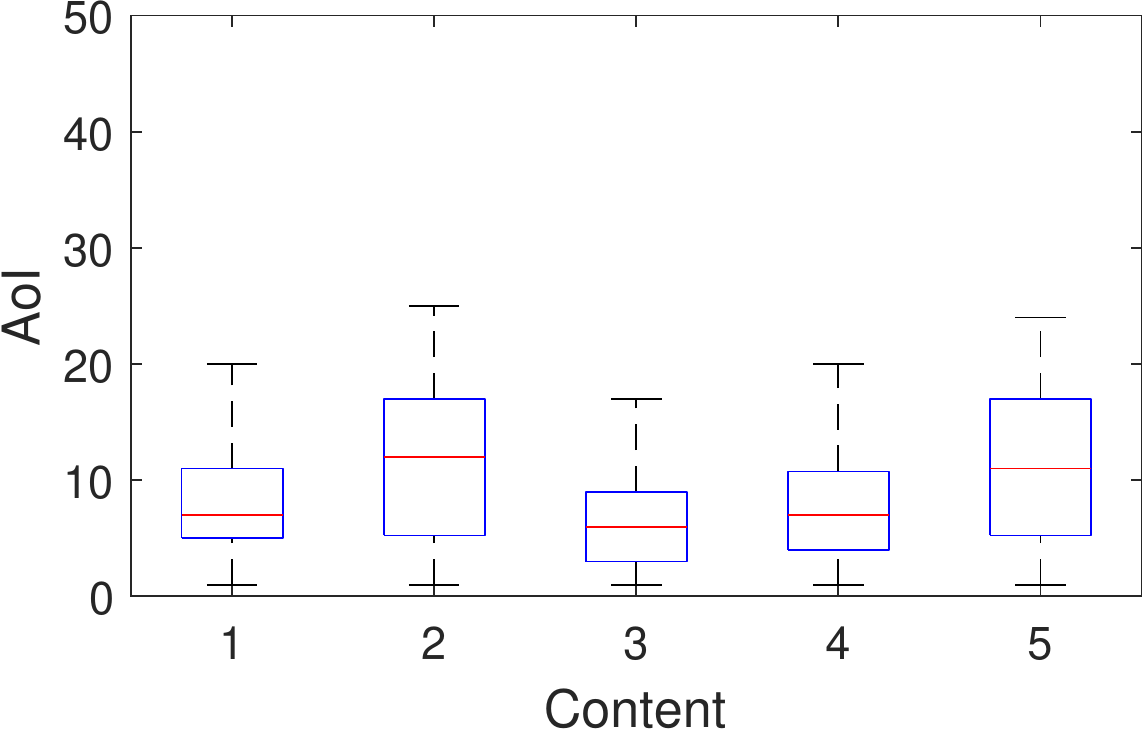}\captionsetup{justification=centering}
     \subcaption{RSU 1 with proposed algorithm}
     \includegraphics[width=0.9\linewidth]{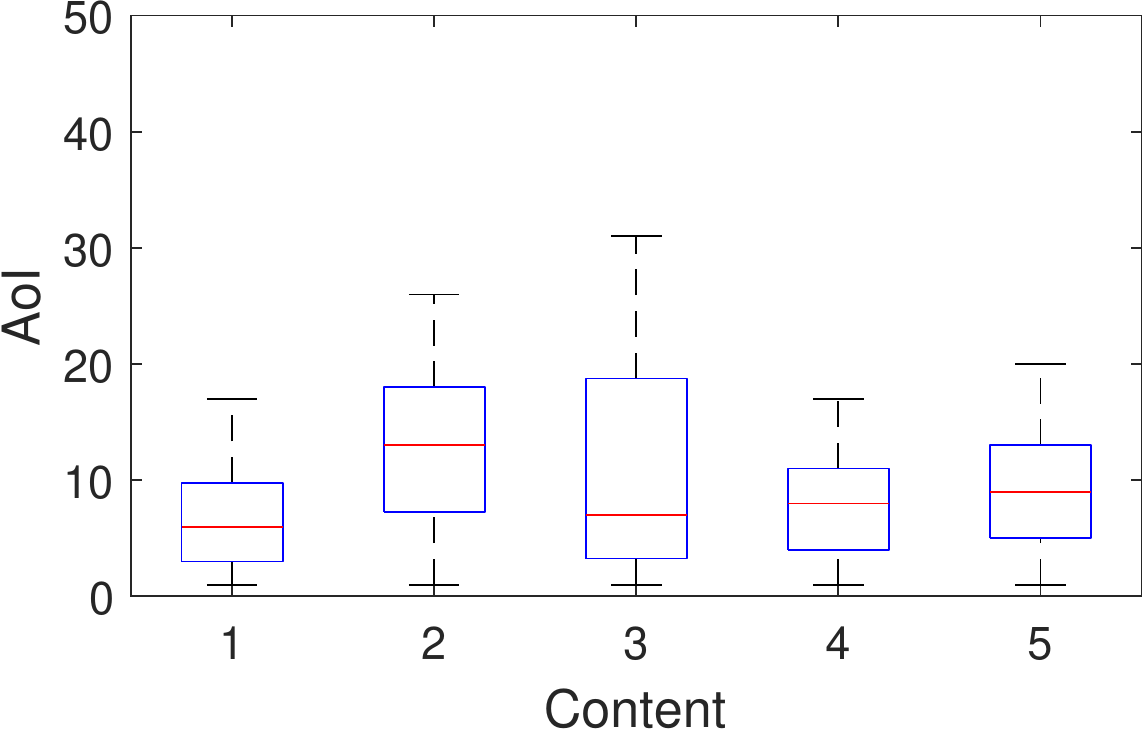}\captionsetup{justification=centering}
     \subcaption{RSU 2 with proposed algorithm}
     \includegraphics[width=0.9\linewidth]{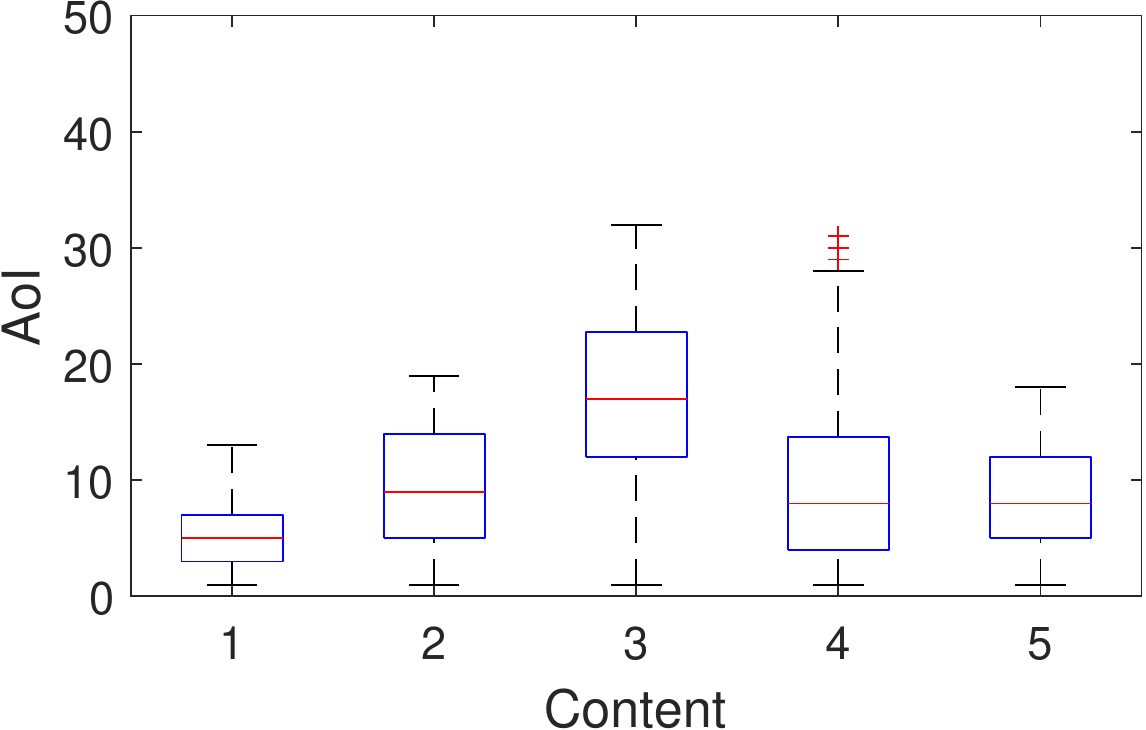}\captionsetup{justification=centering}
     \subcaption{RSU 3 with proposed algorithm}
     \includegraphics[width=0.9\linewidth]{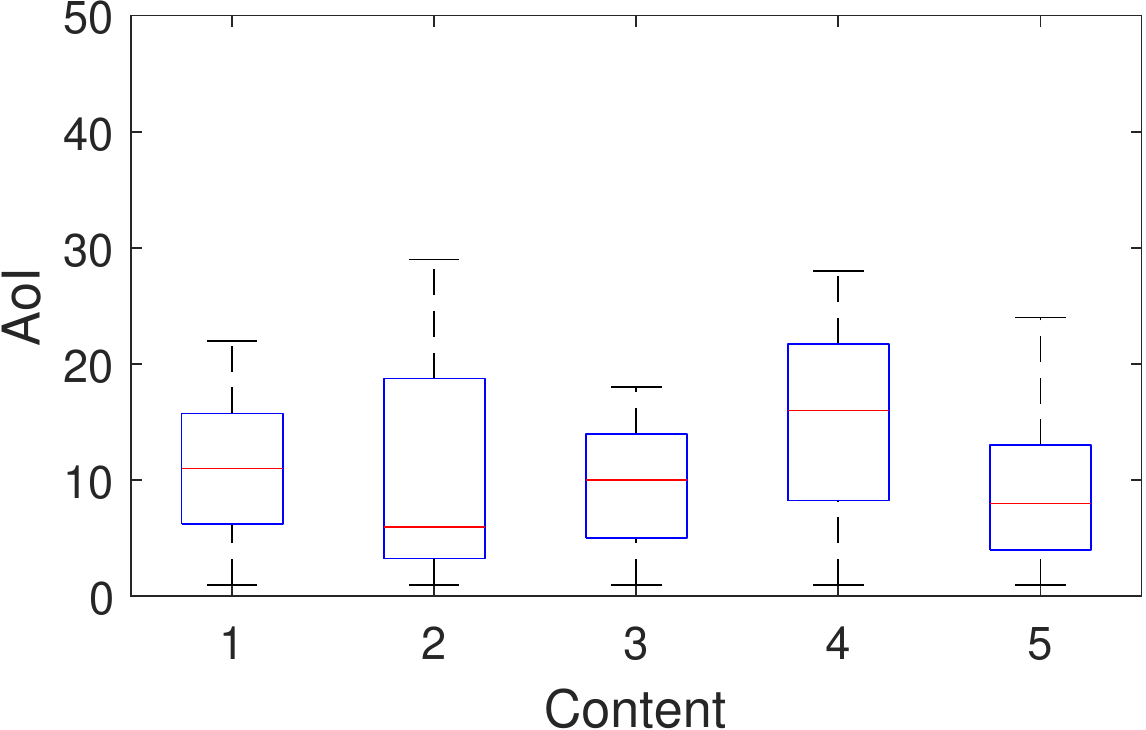}\captionsetup{justification=centering}
     \subcaption{RSU 4 with proposed algorithm}     
    \end{multicols}
    
    \begin{multicols}{4}
     \includegraphics[width=0.9\linewidth]{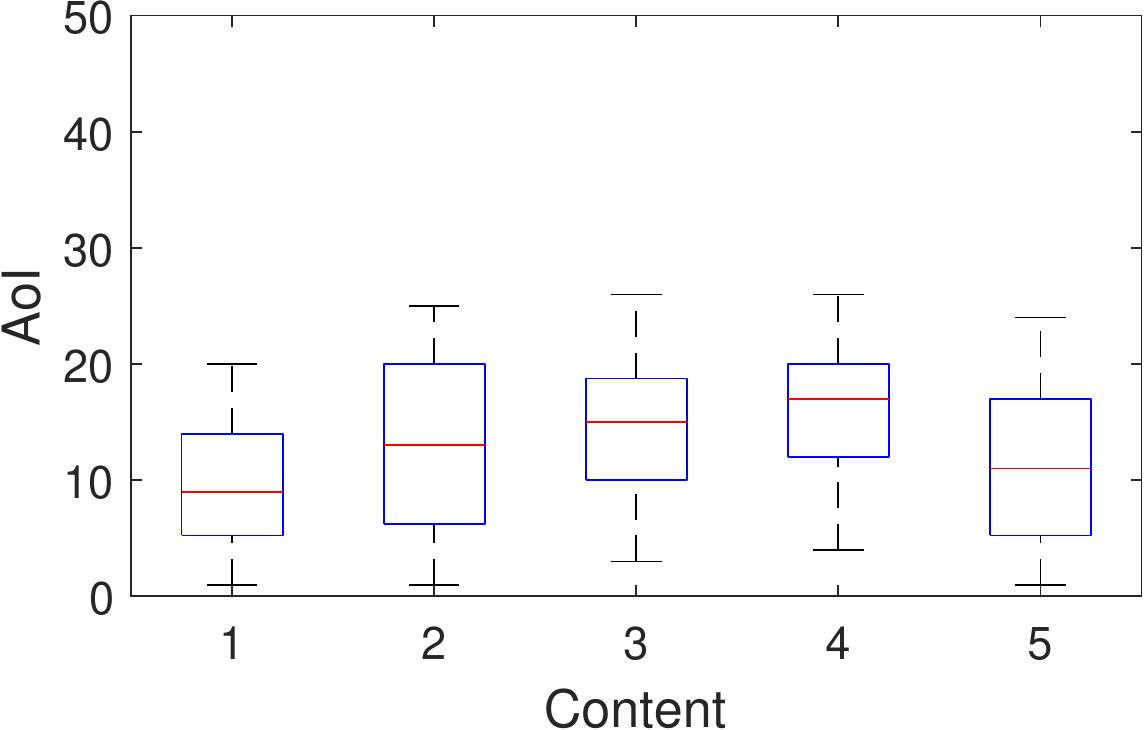}\captionsetup{justification=centering}
     \subcaption{RSU 1 with AoI-greedy algorithm}
     \includegraphics[width=0.9\linewidth]{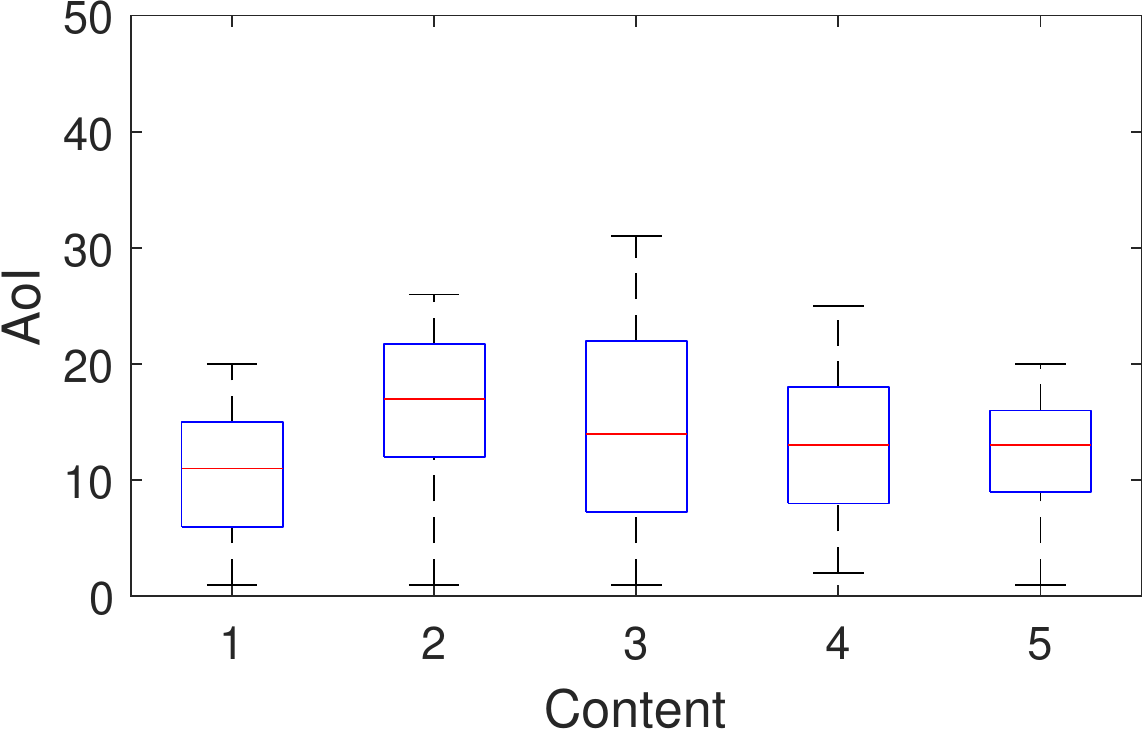}\captionsetup{justification=centering}
     \subcaption{RSU 2 with AoI-greedy algorithm}
     \includegraphics[width=0.9\linewidth]{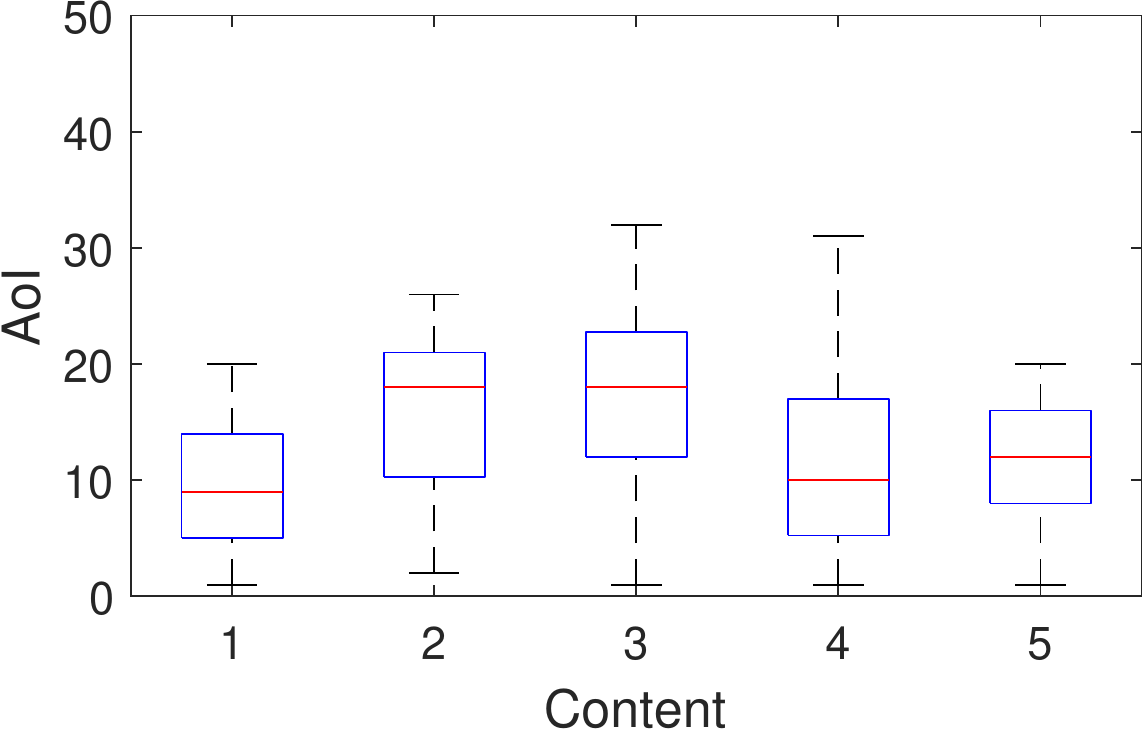}\captionsetup{justification=centering}
     \subcaption{RSU 3 with AoI-greedy algorithm}
     \includegraphics[width=0.9\linewidth]{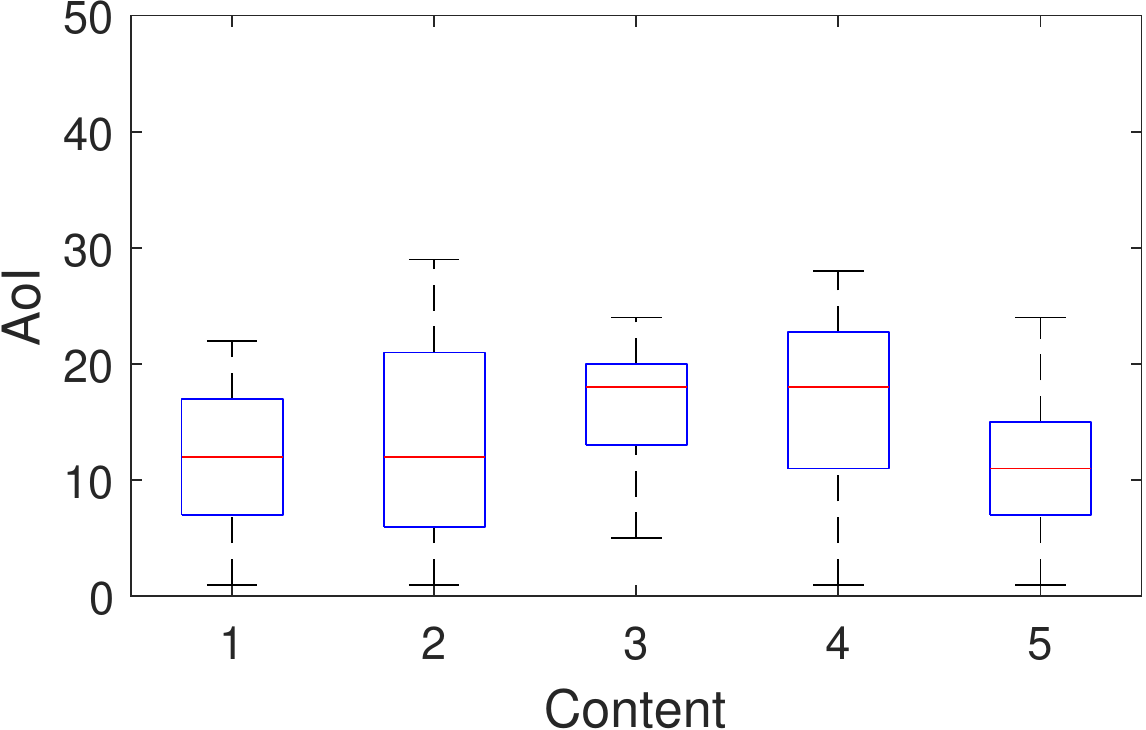}\captionsetup{justification=centering}
     \subcaption{RSU 4 with AoI-greedy algorithm}     
    \end{multicols}
\caption{Average AoI for all regions present in the connected vehicle systems.}%
\label{fig:average AoI for all regions}
\end{figure*}

% region별 average AoI
The average content AoI state of the total $20$ regions can be confirmed in Fig.~\ref{fig:average AoI for all regions}. Fig.~\ref{fig:average AoI for all regions}(a)$\sim$(d) and Fig.~\ref{fig:average AoI for all regions}(e)$\sim$(h) are the results of using proposed and AoI-greedy algorithms under the same conditions, respectively. In the graph, the red solid line means the average value for the $100$-unit time. In the results of the two algorithms, the maximum and minimum values are similar, but for the interquartile range, represented by a solid blue box, the proposed algorithm always has a lower AoI range for all $20$ regions.

% sorted AoI
\begin{figure}
    \begin{multicols}{2}\centering
     \includegraphics[width =0.9\linewidth]{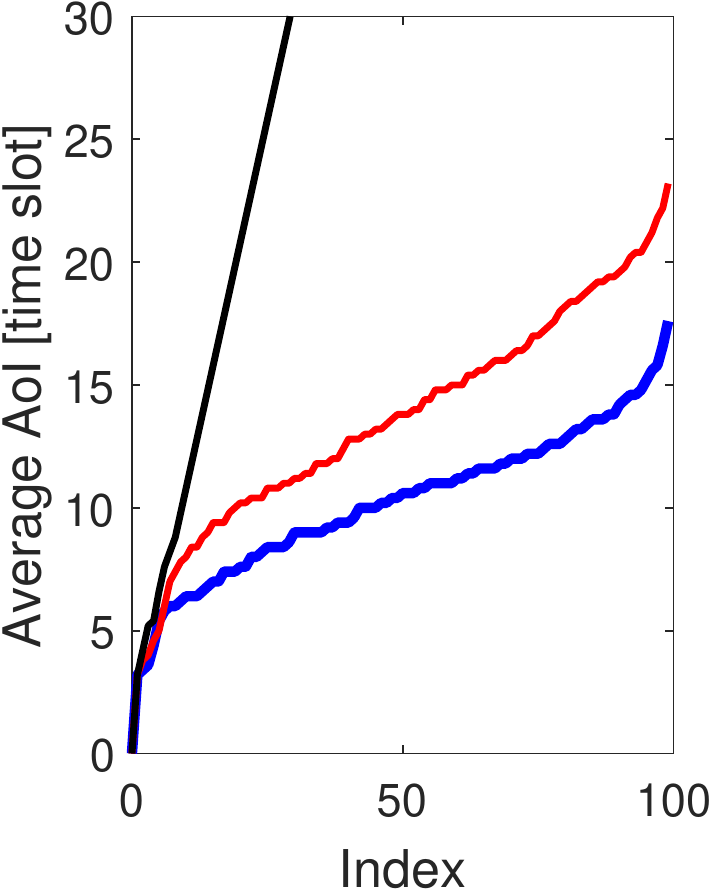}\captionsetup{justification=centering}
     \subcaption{Sorted average AoI value}
     \includegraphics[width =0.9\linewidth]{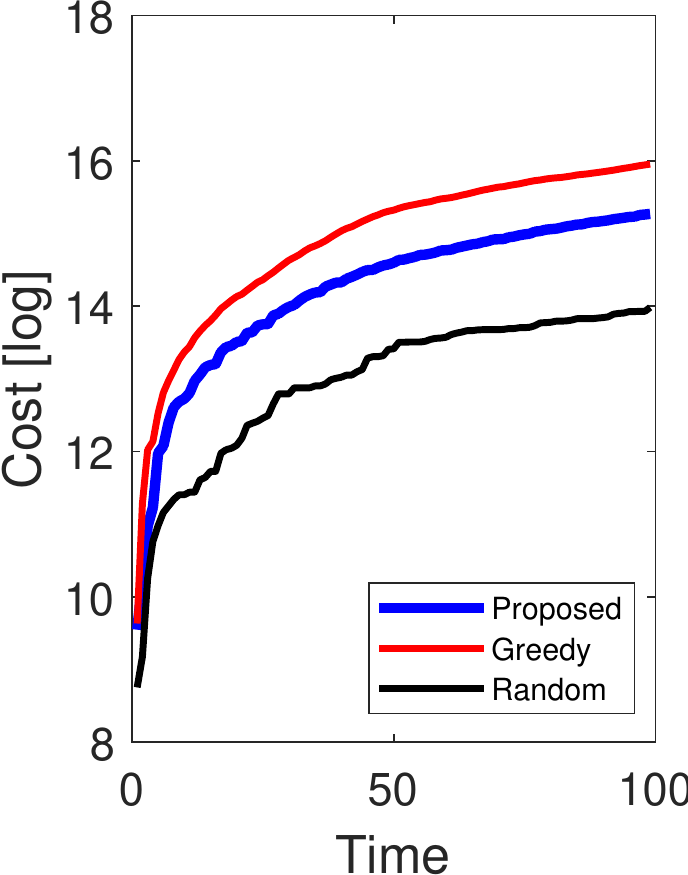}\captionsetup{justification=centering}
     \subcaption{Cost fluctuation over time}
    \end{multicols}
\caption{AoI-aware content caching results.}
\label{fig:sorted AoI}
\end{figure}

% ==================== Table: number of AoI max over ====================
\begin{table}[t]
\centering
\caption{Cumulative number of times the AoI value of the RSU exceeds the AoI max.}
\begin{tabular}{c||ccc}
\toprule[1.0pt]
                & Proposed & AoI-greedy & Random \\     
\midrule[1.0pt]
Updates  & 260       & 297     & 146      \\
$A^{max}_h$ over & 638      & 1018     & 1741     \\     
\bottomrule[1.0pt]
\end{tabular}
\label{tab:AoI max over}
\end{table}

% ==================== Table: AoI minimum and maximum values compared to $AoI^{max}_h$ by region type  ====================
% (I) Max AoI Difference => medium
\begin{table*}[t]\centering
\caption{RSUs' content AoI compared to $AoI^{max}_h$ by region type.}
\label{tab: average AoI value}
%\text{[normal: 20, traffic jam: 10, accident: 8, crowded: 15]}
\subcaption{W/ the proposed algorithm}
\begin{tabular}{c||rrr|rrr|rrr|rrr}
\toprule[1.0pt]
\multirow{2}{*}{time} &
  \multicolumn{3}{c|}{Normal} &
  \multicolumn{3}{c|}{Traffic Jam} &
  \multicolumn{3}{c|}{Accident} &
  \multicolumn{3}{c}{Crowded}\\ \cline{2-13} 
 &
  \multicolumn{1}{c|}{avg} &
  \multicolumn{1}{c|}{min} &
  \multicolumn{1}{c|}{max} &
  \multicolumn{1}{c|}{avg} &
  \multicolumn{1}{c|}{min} &
  \multicolumn{1}{c|}{max} &
  \multicolumn{1}{c|}{avg} &
  \multicolumn{1}{c|}{min} &
  \multicolumn{1}{c|}{max} &
  \multicolumn{1}{c|}{avg} &
  \multicolumn{1}{c|}{min} &
  \multicolumn{1}{c}{max} \\ 
\midrule[1.0pt]
10  & 7.6 & 1 & 13 & 11 & 2 & 15 & 6.8 & 3 & 11 & 8.5 & 3 & 14 \\ \cline{1-1}
20 & 13.6 & 3 & 22 & 12.6 & 3 & 23 & 11.4 & 5 & 18 & 9.3 & 2 & 20 \\ \cline{1-1}
30 & 9.2 & 2 & 22 & 15.2 & 3 & 23 & 8 & 1 & 16 & 11.3 & 6 & 23 \\ \cline{1-1}
40 & 8 & 3 & 12 & 9.8 & 3 & 22 & 13 & 3 & 25 & 11.3 & 1 & 22\\ \cline{1-1}
50 & 10 & 2 & 22 & 11.6 & 3 & 29 & 13.2 & 1 & 25 & 11.8 & 3 & 29\\ \cline{1-1}
60 & 6.4 & 1 & 12 & 17.4 & 12 & 25 & 10 & 3 & 29 & 9.6 & 1 & 22\\ \cline{1-1}
70 & 11.8 & 2 & 22 & 12.8 & 6 & 25 & 8 & 3 & 15 & 5.8 & 1 & 16\\ \cline{1-1}
80 & 6.4 & 3 & 15 & 12.4 & 7 & 21 & 6 & 1 & 13 & 11.3 & 3 & 21\\ \cline{1-1}
90 & 11.8 & 2 & 16 & 9.2 & 3 & 21 & 12.4 & 8 & 18 & 17.6 & 2 & 21\\ \cline{1-1}
100 & 4.6 & 2 & 11 & 10 & 1 & 16 & 13.6 & 6 & 20 & 11.6 & 8 & 16\\
\bottomrule[1.0pt]
\end{tabular}

\bigskip

\subcaption{W/ the AoI-greedy algorithm}
\begin{tabular}{c||rrr|rrr|rrr|rrr}
\toprule[1.0pt]
\multirow{2}{*}{time} &
  \multicolumn{3}{c|}{Normal} &
  \multicolumn{3}{c|}{Traffic Jam} &
  \multicolumn{3}{c|}{Accident} &
  \multicolumn{3}{c}{Crowded}\\ \cline{2-13} 
 &
  \multicolumn{1}{c|}{avg} &
  \multicolumn{1}{c|}{min} &
  \multicolumn{1}{c|}{max} &
  \multicolumn{1}{c|}{avg} &
  \multicolumn{1}{c|}{min} &
  \multicolumn{1}{c|}{max} &
  \multicolumn{1}{c|}{avg} &
  \multicolumn{1}{c|}{min} &
  \multicolumn{1}{c|}{max} &
  \multicolumn{1}{c|}{avg} &
  \multicolumn{1}{c|}{min} &
  \multicolumn{1}{c}{max} \\
\midrule[1.0pt]
10  & 12&11&13&12&11&13&10.2&7&12&11.5&10&22 \\ \cline{1-1}
20 & 15.6& 3&22&18.25&8&23&20.2&17&22&17&2&26 \\ \cline{1-1}
30 & 12.8&3&22&14.5&8&19&8.4&1&18&13.5&8&25 \\ \cline{1-1}
40 &10.4&3&16&8&3&18&14.6&9&25&17.1&8&29\\ \cline{1-1}
50 &19.2&13&23&12.25&5&16&16.2&8&25&13.6&3&31\\ \cline{1-1}
60 & 15.6&1&23&22.2&15&26&18&3&29&16.6&8&29\\ \cline{1-1}
70 & 16.4&11&22&12&6&25&11.6&6&16&8&1&21\\ \cline{1-1}
80 & 7.8&3&15&15.2&8&19&9.8&1&18&13&9&24\\ \cline{1-1}
90 & 11.8&2&16&14.5&5&18&13.2&9&18&9.6&2&21\\ \cline{1-1}
100 & 5.6&2&11&10.5&1&15&14.4&6&21&13.5&9&17\\
\bottomrule[1.0pt]
\end{tabular}
\end{table*}

Fig.~\ref{fig:sorted AoI} represents performance of the proposed algorithm with the comparison target methods over time. In each sub figure, Fig.~\ref{fig:sorted AoI}(a) shows the sorted AoI values for normal type regions existing in the system during $100$ times using the proposed, AoI-greedy, and random methods. There are characteristics that the average AoI value of proposed algorithm maintains the smallest state for the most number of times. When comparing the proposed and AoI gready algorithms considering AoI, except for random algorithms that update only randomly selected regions among the total $20$ contents, since AoI-greedy method ignores the concept of the maximum effective time of the contents, $A^{max}_h$ AoI-greedy method is mainly distributed in values slightly larger than the proposed algorithm. The maximum AoI value of each caching algorithms during the entire time is distributed between $10\sim15$ and $15\sim20$, respectively. Theses results show that the proposed caching algorithm manages valid content for each region type by considering the maximum validity compared to other comparative algorithms. 
Fig.~\ref{fig:sorted AoI}(b) represents the log scale CDF of communication cost which occurs during content caching. Each algorithm is in the same form in which the graph increases. However, there is a difference in the increase in the cost value. Although random with little content upload and update via MBS consumes the smallest cost, comparing the two algorithms (e.g., proposed and AoI-greedy) that normally manage RSU cache, Fig.~\ref{fig:sorted AoI}(b) shows that the proposed algorithm best reflects the latest situation on road content while using less communication cost.
Tab.~\ref{tab:AoI max over} and Fig.~\ref{fig:average AoI for all regions} also represent that the proposed algorithm performs better than the others.
Tab.~\ref{tab:AoI max over} shows the number of communication (updates) between the MBS and RSUs that occurred for the management and cumulative number of times greater than $A^{max}_h$ for the total $20$ contents present in the system during $100$-unit time.

% region type별 average AoI

We can check the results of the proposed algorithm in more detail with Tab.~\ref{tab: average AoI value}(a) and Tab.~\ref{tab: average AoI value}(b). The two tables show the average, maximum, and minimum AoI values for each region type every $10$ times interval when the max AoI values ($A^{max}_h$) according to the region type are in case using proposed and AoI-greedy algorithms. These results explain that the proposed algorithm maintains more optimal average AoI state considering $A^{max}_h$ of each type than the order one.

\subsubsection{Performance of the Request Service Algorithm}

\begin{figure*}[t!]\centering
 \begin{multicols}{3}
     \includegraphics[width=\linewidth]{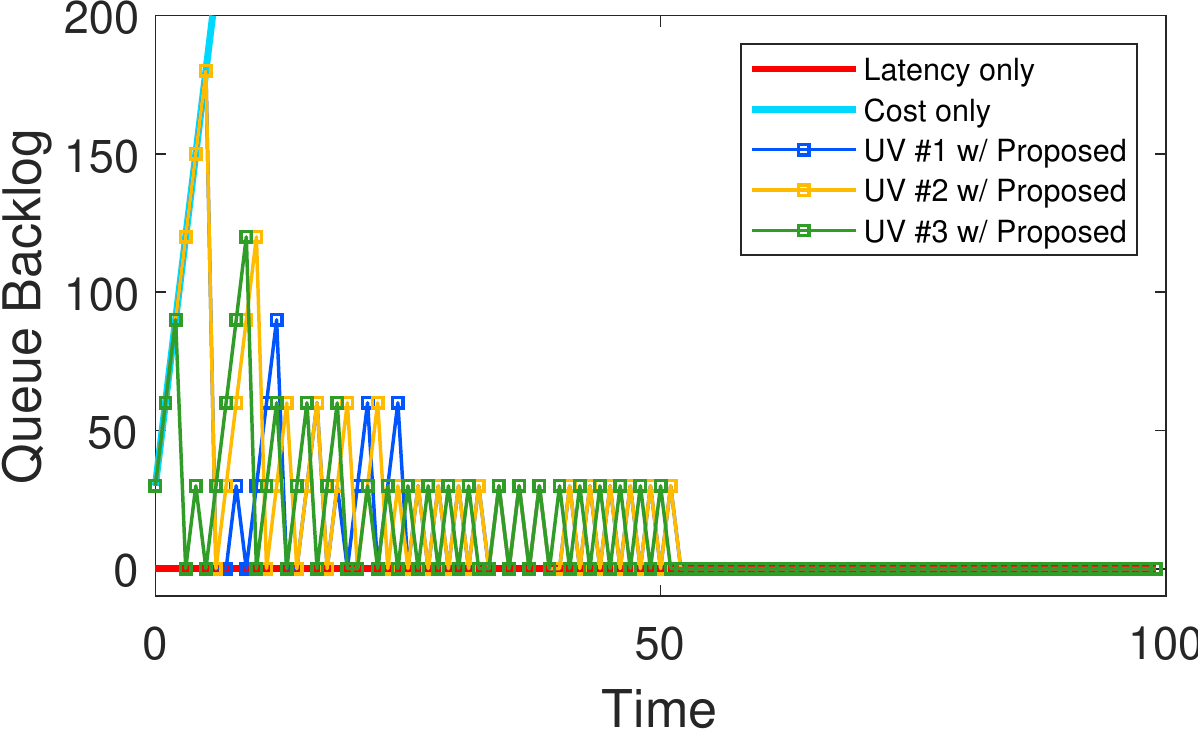}\captionsetup{justification=centering}
     \subcaption{Light weight value}\label{fig:light_q}
     \includegraphics[width=\linewidth]{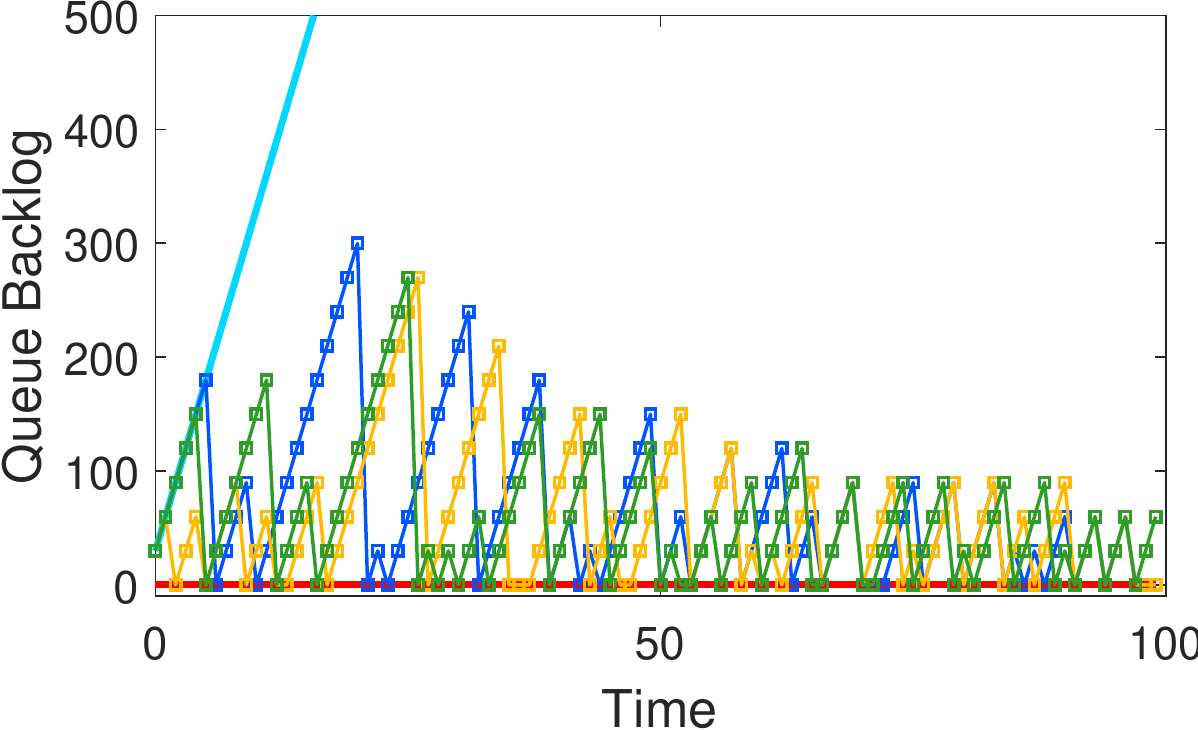}\captionsetup{justification=centering}
     \subcaption{Normal weight value}\label{fig:normal_q}
     \includegraphics[width=\linewidth]{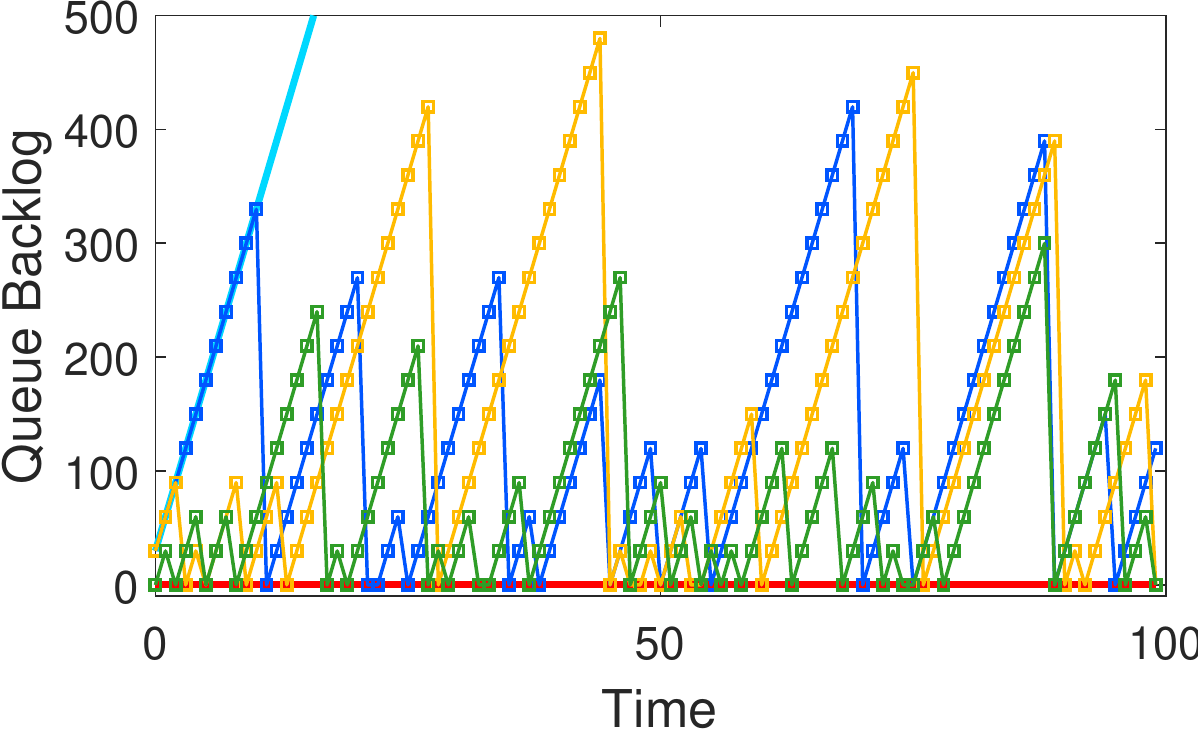}\captionsetup{justification=centering}
     \subcaption{Heavy weight value}\label{fig:heavy_q}
 \end{multicols}
 \begin{multicols}{3}
     \includegraphics[width=\linewidth]{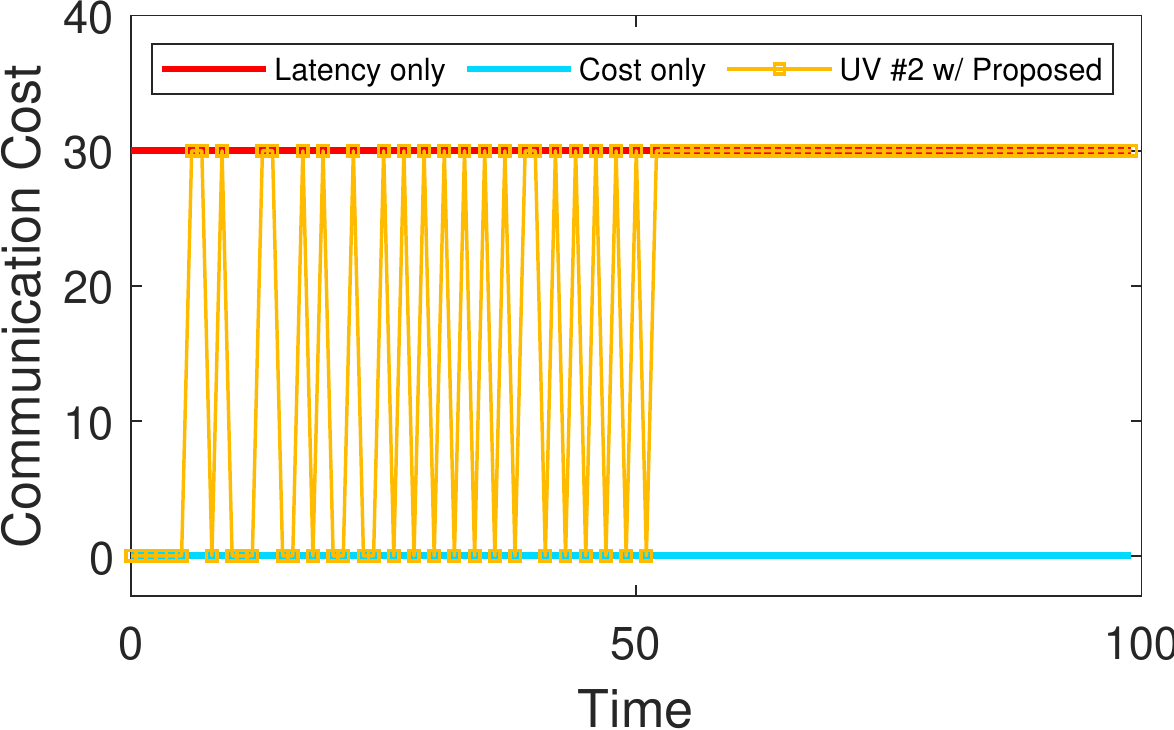}\captionsetup{justification=centering}
     \subcaption{Light weight value}\label{fig:light_c}
     \includegraphics[width=\linewidth]{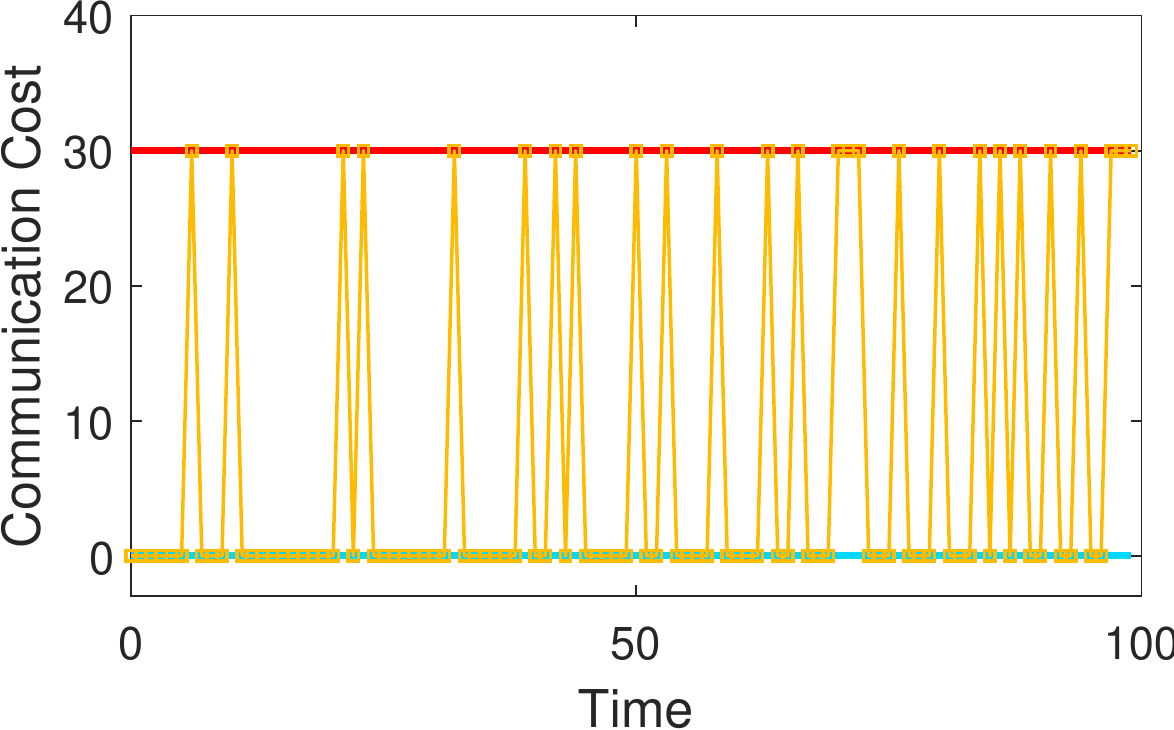}\captionsetup{justification=centering}
     \subcaption{Normal weight value}\label{fig:normal_c}
     \includegraphics[width=\linewidth]{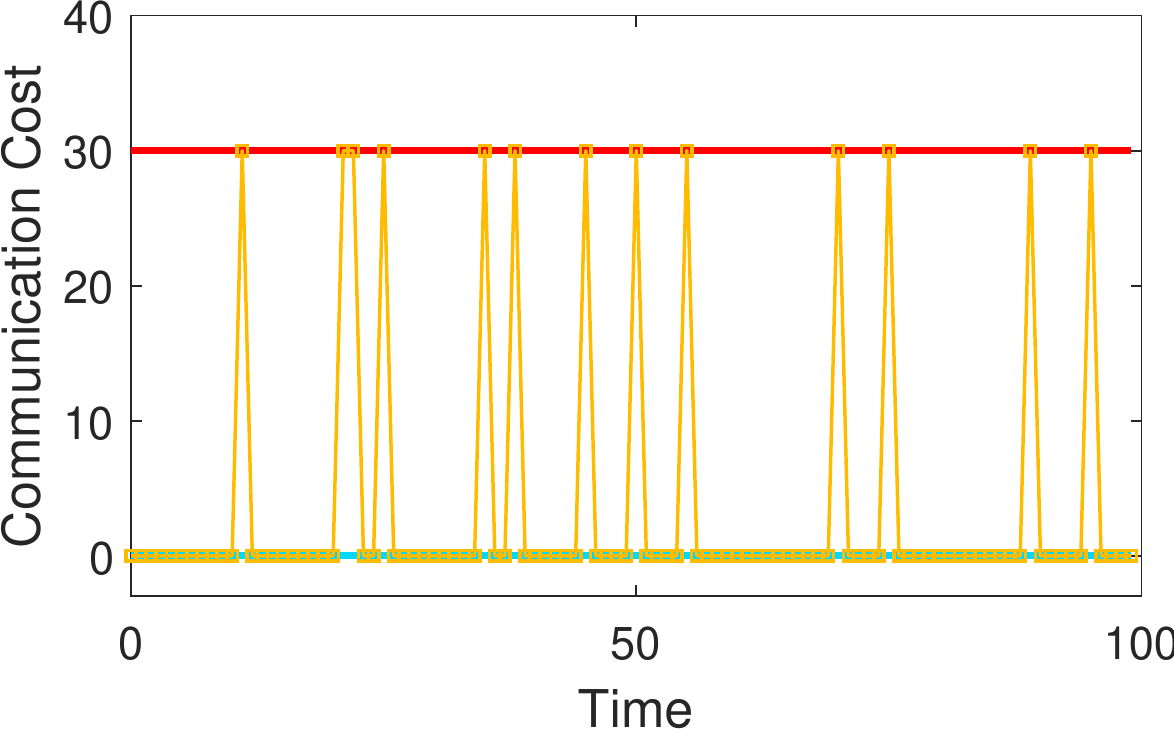}\captionsetup{justification=centering}
     \subcaption{Heavy weight value}\label{fig:heavy_c}
 \end{multicols}
 \caption{Delay-aware content service according to the weight value $V$ variation.}%
 \label{fig:Lyapunov simul}%
 \vspace{-3mm}
\end{figure*}

In this part, we turn to evaluating the performance of Lyapunov optimization based UV request service control. For the optimal service controlling performance evaluation, we adopt the two types of comparative strategies. One is a method only considering communication cost minimization, and the other is only for the waiting queue stability (latency).
We show that our proposed algorithm is excellent by using the two extreme methods as a base line.
In Fig.~\ref{fig:Lyapunov simul}, various experimental results according to the different value of $V$, which means where the weight is placed on the communication cost and queue stability in the trade-off relationship, are described.

The algorithm is applied in the presence of one RSU, $3$ UVs, and $5$ regions.
% queue backlog 설명
Each UV requests the RSU for the targeted region content, and the RSU must complete the service before each UV passes the target region (i.e., before the waiting queue is expired). The queue backlog for each UV is accumulated until receiving the requested service, which means that the linear increase in Fig.~\ref{fig:Lyapunov simul}(a), Fig.~\ref{fig:Lyapunov simul}(b), and Fig.~\ref{fig:Lyapunov simul}(c).
If the service is provided by the proposed algorithm, the queue backlog is cleaned according to the~\eqref{eq:lyapunov-final}, and the queue backlog is repeatedly accumulated again after the UV sends a new service request for another target region.
The two methods (e.g., latency only, cost only) that are the comparison methods of the proposed algorithm are represented by red and cyan solid lines, respectively. Each one means (i) RSU always services all of the requests immediately, regardless of the amount of communication cost, and (ii) RSU does not service even if the latency (queue backlog) of each UV is exceeded to minimize the communication cost.
It can be seen that the queue backlog result differs depending on the size of the $V$ value.
The larger the $V$, the more weighted the cost minimization, which is the object function~\eqref{eq:opt}.
For this reason, the queue backlog upper limit in the light weight case has a smaller value than in the normal weight case, and on the contrary, in the heavy weight case, the upper limit is the largest among the three cases.
That is, the heavier the weight case, the longer the waiting time for the UV to wait for the service may be allowed.
However, as expressed in~\eqref{eq:queue validity }, if the delay is accumulated until the UV which requests the content passes the target region, it leads to failure to actually serve the UV even if the backlog of the waiting queue is not overflowing.
For this reason, it is important to find an appropriate value of $V$ and to control the waiting delay and communication cost so that each RSU can service UVs before passing the target regions.
This can also be seen in Tab.~\ref{tab:lyapunov}. There is a difference in the ratio of the number of service waits and the number of service successes according to the $V$ value within the same time.
\begin{table}[t]
\centering
\caption{The comparison of content service completion figures according to the weight value $V$ in~\eqref{eq:lyapunov-final}.}
\begin{tabular}{c||ccc}
\toprule[1.0pt]
                & Normal $V$ & Light $V$ & Heavy $V$ \\ 
\midrule[1.0pt]
service success & 51       & 151     & 38      \\
cost save       & 245      & 141     & 257     \\ 
\bottomrule[1.0pt]
\end{tabular}
\label{tab:lyapunov}
\end{table}
% cost 설명 + table
In Fig.~\ref{fig:Lyapunov simul}(d), Fig.~\ref{fig:Lyapunov simul}(e), and Fig.~\ref{fig:Lyapunov simul}(f), we can check that the proposed algorithm and two comparison methods communication cost during $100$ unit-time. 
As described above, latency only and cost only methods are represented as red and cyan solid lines, respectively, and always record the maximum cost and $0$ cost.
According to the proposed algorithm, when supporting the requests of the $2$nd UV, depending on the value of $V$,
light weight case allows shorter latency and enables more service support success. Heavy weight case can reduce total communication costs by supporting the least number of services with longer latency during $100$ unit-times.

\section{Concluding Remarks}\label{sec:conclusion}

This paper proposed a two-stage joint AoI-aware cache management and content delivery scheme for providing fresh road contents for connected vehicles. 
Optimization of content transmission decisions for the distributed cache management considering the concept of data freshness in the system and content service using road infrastructure are essential. Therefore, a new dynamic decision algorithms based on Markov Decision Process (MDP) and Lyapunov optimization applying AoI is important.
We present the MDP-based algorithm for cache management of RSUs to limit the content AoI of cached contents as relatively up to date.
In addition, the content delivery from cache-enabled RSUs to UVs which is adaptively optimized depending on the current AoI of contents and rapidly time-varying traffic conditions under the Lyapunov-based control is also proposed.
The proposed scheme adaptively controls the trade-off between the content AoI and network resource consumption, depending on rapidly changing road environments, user mobility, as well as the AoI of contents. Furthermore, the performance of the proposed research technology is verified through various experiments.

\bibliographystyle{IEEEtran}
% \bibliography{ref_mobilebs,ref_aimlab,ref_aoi}

\begin{IEEEbiography}[{\includegraphics[width=1in,height=1.25in,clip,keepaspectratio]{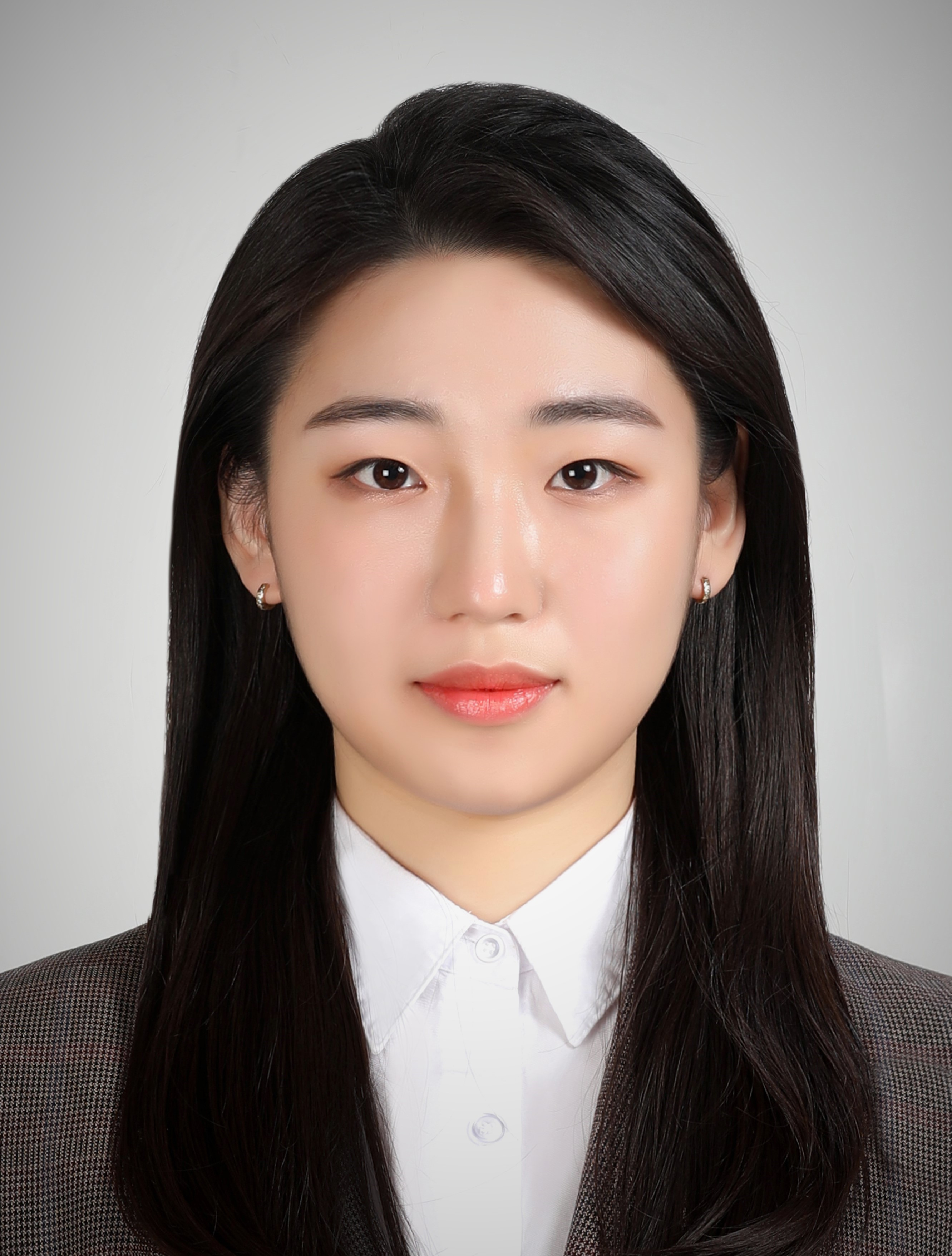}}]{Soohyun Park} is currently pursuing the Ph.D. degree in electrical and computer engineering at Korea University, Seoul, Republic of Korea. She received the B.S. degree in computer science and engineering from Chung-Ang University, Seoul, Republic of Korea, in 2019. Her research focuses include deep learning algorithms and their applications to big-data platforms and networking. 

She was a recipient of the IEEE Vehicular Technology Society (VTS) Seoul Chapter Award in 2019.\end{IEEEbiography}

\begin{IEEEbiography}[{\includegraphics[width=1in,height=1.25in,clip,keepaspectratio]{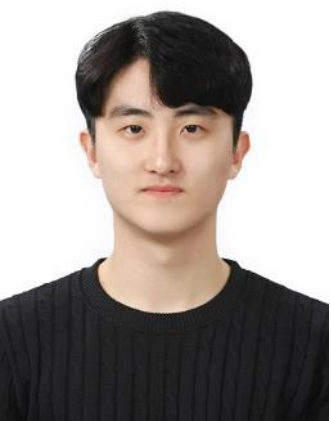}}]{Chanyoung Park} is currently a Ph.D. student in Electrical and Computer Engineering at Korea University, Seoul, Republic of Korea, since September 2022. He received the B.S. degree in electrical and computer engineering from Ajou University, Suwon, Republic of Korea, in 2022, with honor (early graduation). His research focuses include deep learning algorithms and their applications to networks. 
\end{IEEEbiography}

\begin{IEEEbiography}[{\includegraphics[width=1in,height=1.25in,clip,keepaspectratio]{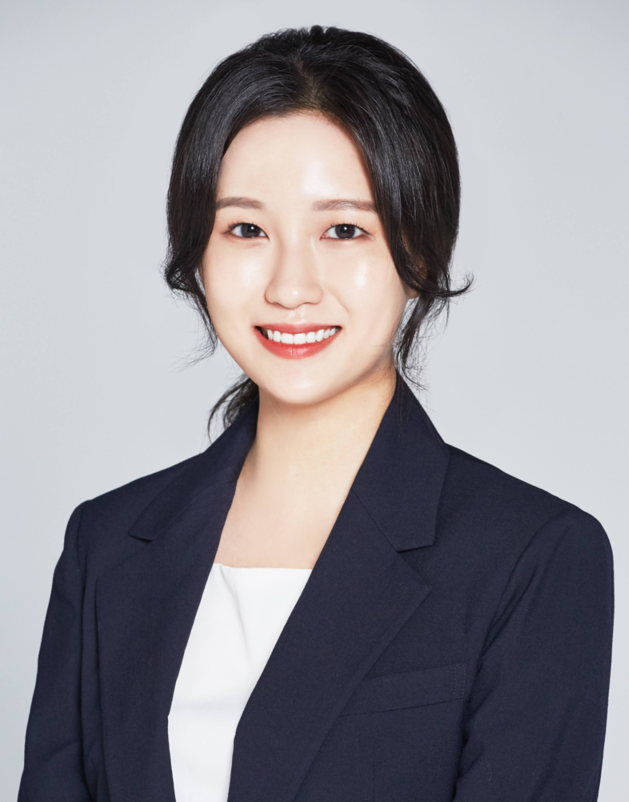}}]{Soyi Jung} has been an assistant professor at the Department of Electrical of Computer Engineering, Ajou University, Suwon, Republic of Korea, since September 2022. Before joining Ajou University, she was an assistant professor at Hallym University, Chuncheon, Republic of  Korea, from 2021 to 2022; a visiting scholar at Donald Bren School of Information and Computer Sciences, University of California, Irvine, CA, USA, from 2021 to 2022; a research professor at Korea University, Seoul, Republic of Korea, in 2021; and a researcher at Korea Testing and Research (KTR) Institute, Gwacheon, Republic of Korea, from 2015 to 2016. She received her B.S., M.S., and Ph.D. degrees in electrical and computer engineering from Ajou University, Suwon, Republic of Korea, in 2013, 2015, and 2021, respectively. Her current research interests include network optimization for autonomous vehicles communications, distributed system analysis, big-data processing platforms, and probabilistic access analysis. She was a recipient of Best Paper Award by KICS (2015), Young Women Researcher Award by WISET and KICS (2015), Bronze Paper Award from IEEE Seoul Section Student Paper Contest (2018), ICT Paper Contest Award by Electronic Times (2019), and IEEE ICOIN Best Paper Award (2021).
\end{IEEEbiography}

\begin{IEEEbiography}[{\includegraphics[width=1in,height=1.25in,clip,keepaspectratio]{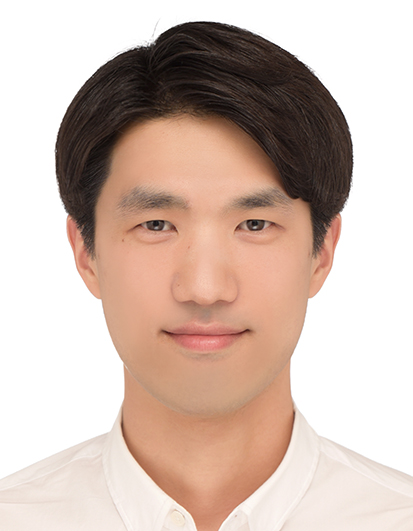}}]{Minseok Choi} is an Assistant Professor in Electronic Engineering with Kyung Hee University, Yongin, South Korea. He received the B.S., M.S., and Ph.D. degrees from the School of Electrical Engineering, Korea Advanced Institute of Science and Technology (KAIST), Daejeon, South Korea, in 2011, 2013, and 2018, respectively. He was an Assistant Professor in telecommunications engineering with Jeju National University, a Visiting Postdoctoral Researcher in electrical and computer engineering with the University of Southern California (USC), Los Angeles, CA, USA, and a Research Professor in electrical engineering with Korea University, Seoul, South Korea. He received the IEEE Communications Society (ComSoc) Multimedia Communications Technical Committee (MMTC) Best Paper Award, 2022. His research interests include wireless caching networks, federated learning, stochastic network optimization, wireless intelligent networks.
\end{IEEEbiography}

\begin{IEEEbiography}[{\includegraphics[width=1in,height=1.25in,clip,keepaspectratio]{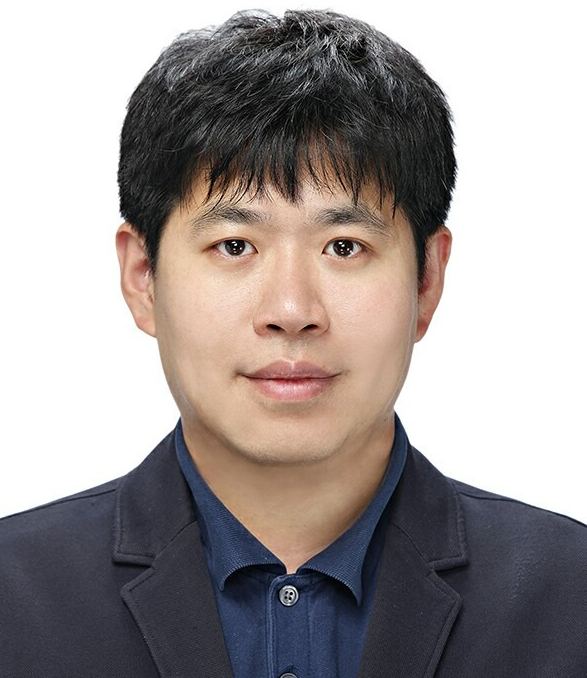}}]{Joongheon Kim}
(M'06--SM'18) has been with Korea University, Seoul, Korea, since 2019, and he is currently an associate professor. He received the B.S. and M.S. degrees in Computer Science and Engineering from Korea University, Seoul, Korea, in 2004 and 2006, respectively; and the Ph.D. degree in Computer Science from the University of Southern California (USC), Los Angeles, CA, USA, in 2014. Before joining Korea University, he was with LG Electronics (Seoul, Korea, 2006--2009), InterDigital (San Diego, CA, USA, 2012), Intel Corporation (Santa Clara in Silicon Valley, CA, USA, 2013--2016), and Chung-Ang University (Seoul, Korea, 2016--2019). 

He is a senior member of the IEEE, and serves as an associate editor for \textsc{IEEE Transactions on Vehicular Technology}.

He was a recipient of Annenberg Graduate Fellowship with his Ph.D. admission from USC (2009), 
Intel Corporation Next Generation and Standards (NGS) Division Recognition Award (2015), Haedong Young Scholar Award by KICS (2018), IEEE Vehicular Technology Society (VTS) Seoul Chapter Award (2019), Outstanding Contribution Award by KICS (2019), Gold Paper Award from IEEE Seoul Section Student Paper Contest (2019), Granite Tower Best Teaching Award by Korea University (2020), \textsc{IEEE Systems Journal} Best Paper Award (2020), and IEEE ICOIN Best Paper Award (2021).
\end{IEEEbiography}
\end{document}